\documentclass[%
 reprint,
superscriptaddress,
nofootinbib,
 amsmath,amssymb,
 aps,
]{revtex4-1}

\usepackage{graphicx}
\usepackage{dcolumn}
\usepackage{bm}
\usepackage{xcolor}

\usepackage{placeins}
\usepackage{subfigure}
\usepackage{supertabular}

\newcommand{\fgw}{f_\mathrm{gw}}
\newcommand{\dotfgw}{\dot{f}_{\rm gw}}
\newcommand{\fgwini}{f_\mathrm{gw0}}
\newcommand{\dotfgwini}{\dot{f}_{\rm gw0}}

\graphicspath{{./fig/}}

\begin{document}

\preprint{APS/123-QED}

\title{Application of hidden Markov model tracking to the search for long-duration transient gravitational waves from the remnant of the binary neutron star merger GW170817}

\author{Ling Sun}
\email[]{lssun@caltech.edu}
\affiliation{LIGO Laboratory, California Institute of Technology, Pasadena, California 91125, USA}
\affiliation{School of Physics, University of Melbourne, Parkville, Victoria 3010, Australia}
\affiliation{Australian Research Council Centre of Excellence for Gravitational Wave Discovery (OzGrav)}

\author{Andrew Melatos}
\affiliation{School of Physics, University of Melbourne, Parkville, Victoria 3010, Australia}
\affiliation{Australian Research Council Centre of Excellence for Gravitational Wave Discovery (OzGrav)}

\date{\today}

\begin{abstract}


The nature of the post-merger remnant of GW170817, the first binary neutron star coalescence observed by the Advanced Laser Interferometer Gravitational Wave Observatory (Advanced LIGO) and Advanced Virgo, is unknown. Searches have been carried out for short ($\lesssim 1$\,s), intermediate ($\lesssim 500$\,s), and long ($\sim$ days) signals using various algorithms without yielding a detection. We describe an efficient frequency tracking scheme based on a hidden Markov model to search for long-duration transient signals from a neutron star remnant with spin-down time-scale in the range $\sim 10^2\,{\rm s}$--$10^4$\,s. The method was used to the search for a signal from GW170817. We validate the method and estimate its sensitivity through Monte-Carlo simulations on the same data set as used in the GW170817 search. We describe the search configuration and follow-up procedure step by step. The search achieves an astrophysical reach of $\sim 1$\,Mpc and hence cannot detect a source like GW170817 ($40^{+8}_{-14}$\,Mpc), given the current sensitivities of Advanced LIGO and Advanced Virgo. The methodology of the hidden Markov model is described fully to ensure that future analyses of this kind can be reproduced by an independent party.

\end{abstract}

\maketitle


\section{Introduction}

On August 17, 2017, the Advanced Laser Interferometer Gravitational Wave Observatory (Advanced LIGO) and Advanced Virgo detectors observed their first binary neutron star coalescence (GW170817) at a luminosity distance of $40^{+8}_{-14}$\,Mpc and localized within a 90\% credible sky region of 16\,deg$^2$ \cite{2017-GW170817,Abbott:2018wiz}. The initial masses of the two components in the binary are inferred to lie between $1.00M_\odot$ and $1.89M_\odot$, consistent with the masses of known neutron stars. The total mass of the system is measured to be $2.73^{+0.04}_{-0.01} M_\odot$ \cite{Abbott:2018wiz}. This gravitational-wave event was followed by a short gamma-ray burst (GRB 170817A) observed by the Fermi Gamma-ray Burst Monitor $\sim 1.7$\,s later at the same sky location, providing the first direct evidence that binary neutron star mergers are associated with short gamma-ray bursts \cite{170817GRB}. Subsequent observations of X-ray, ultraviolet, optical, infrared and radio counterparts support the hypothesis that this event was produced by a binary neutron star merger in the galaxy NGC 4993 \cite{170817MMA,170817kilonova}. 

The stellar remnant of GW170817 remains unknown. There are four likely possibilities: (1) a promptly formed black hole, (2) a hypermassive neutron star collapsing into a black hole within $\sim 1$\,s, (3) a supermassive neutron star collapsing into a black hole within $\sim 10$--$10^4$\,s, and (4) a stable neutron star \cite{postmerger2017}. Searches were conducted for short-duration ($\lesssim 1$\,s) gravitational-wave signals using the Coherent Wave Burst (cWB) algorithm \cite{Klimenko2016cwb}, and for intermediate-duration ($\lesssim 500$\,s) signals using both cWB and the Stochastic Transient Analysis Multi-detector Pipeline (STAMP) \cite{Thrane2011-stamp,Thrane:2013bea,Thrane:2014bma}. No signal is detected. The best 50\% detection efficiency upper limits on the root-sum-square strain amplitude are reported as $h_{\rm rss}^{50\%} / (10^{-22} {\rm Hz}^{-1/2})= 2.1$, 8.4, and 5.9 for unmodeled short-duration signals, intermediate-duration millisecond magnetar signals, and intermediate-duration bar-mode signals, respectively \cite{postmerger2017}. An independent short-duration analysis using the BayesWave algorithm \cite{Cornish:2014kda}, which models the post-merger signal from a hypermassive neutron star as a superposition of wavelets, searched 1\,s of data around the coalescence time and placed upper limits on strain amplitude $\sim 3.5$--15 times higher than analytic expectations derived from simulations using different equations of state with source parameters determined from the pre-merger analysis. The strain amplitude upper limits correspond to radiated energy about 12--215 times larger than expectations \cite{Abbott:2018wiz}. 

Searches for long-duration signals were also carried out in 8.5\,days of data from the coalescence to the end of the second observing run (O2). Four pipelines, STAMP \cite{Thrane2011-stamp,Thrane:2013bea,Thrane:2014bma}, Hidden Markov Model (HMM) tracking \cite{Suvorova2016,Sun2018}, Adaptive Transient Hough \cite{Oliver:2018dpt,Krishnan:2004sv,T070124}, and FrequencyHough \cite{Miller:2018genfreqhough,Palomba:2005fp,Antonucci:2008jp,Astone:2014esa}, participated in this analysis. Together they yield a 90\% confidence upper limit on energy radiated in gravitational waves of $\sim 8 M_\odot c^2$ for GW170817 at the measured distance of 40\,Mpc, well above what is plausibly emitted by such a source.
In their current state, the four pipelines have an astrophysical reach of $\sim 1$\,Mpc for events whose gravitational-wave luminosities are comparable to GW170817 \cite{long-duration-pmr}.

HMM tracking provides a computationally efficient strategy for detecting and estimating a quasimonochromatic continuous gravitational wave signal, whose frequency is unknown and evolves due to secular stellar braking and stochastic timing noise \cite{Suvorova2016,Sun2018}. A HMM was applied to data from the first observing run of Advanced LIGO to search for continuous waves from the brightest low-mass X-ray binary, Scorpius X-1, tracking spin wandering caused by fluctuations in the accretion torque \cite{ScoX1ViterbiO1}. It yields a 95\% confidence frequentist upper limit on strain amplitude of $h_0^{95\%} = 4.0 \times 10^{-25}$ at 106 Hz, assuming an electromagnetically-restricted source orientation. A modified HMM that simultaneously tracks secular stellar braking and stochastic timing noise was developed in Ref. \cite{Sun2018} for young supernova remnant searches. The latter algorithm is well suited to searching for a long-transient, quasimonochromatic signal from a binary neutron star post-merger remnant, like GW170817 \cite{2017-GW170817}, if the spin-down time-scale is in the range $10^2\,{\rm s} \lesssim \tau \lesssim 10^4$\,s. In this paper, we describe fully the methodology of a HMM-based post-merger remnant search in order to ensure that future analyses of this kind can be reproduced by independent parties.\footnote{Code and simulation scripts can be found in https://git.ligo.org/ (upon request).} The results from a run on the first binary neutron star coalescence ever observed, GW170817, are published in Ref.~\cite{long-duration-pmr}. For the convenience of the reader, we reproduce the HMM results from Ref.~\cite{long-duration-pmr} in this paper.

The structure of the paper is as follows. In Sec.~\ref{sec:hmm}, we describe the modifications made to the existing HMM method in order to search for long-duration transient signals. In Sec.~\ref{sec:implementation}, we define the detection statistic, discuss the search configuration, and conduct Monte-Carlo simulations to calculate detection threshold and estimate sensitivity. In Sec.~\ref{sec:gw170817}, we describe the data set, parameter space, and set-up of the GW170817 search. Details of the results, follow-up studies, and upper limits (presented in Ref.~\cite{long-duration-pmr}) are provided. A summary of the conclusions is given in Sec.~\ref{sec:conclusion}.

\section{Revised HMM} 
\label{sec:hmm}
\subsection{HMM formulation}
\label{sec:formulation}
A HMM is a memoryless automaton composed of a hidden state variable $q(t) \in \{q_1, \cdots, q_{N_Q}\}$ and measurement variable $o(t)\in \{o_1, \cdots, o_{N_O}\}$ sampled at time $t \in \{t_0, \cdots, t_{N_T}\}$. The most probable sequence of hidden states given the observations over total observing time $T_{\rm obs}$ is computed by the classic Viterbi algorithm \cite{Viterbi1967}. A full description can be found in Refs. \cite{Suvorova2016} and \cite{Sun2018}. In this section, we briefly review the HMM formulation in \citet{Sun2018} and revise the HMM to search for post-merger gravitational waves from GW170817.

Let $\fgw(t)$ be the gravitational-wave frequency at time $t$. We track $q(t)=\fgw(t)$. The discrete hidden states are mapped one-to-one to the frequency bins in the output of a frequency-domain estimator $G(f)$ (defined below) computed over an interval of length $T_{\rm drift}$, with bin size $\Delta f$. We choose $T_{\rm drift}$ to satisfy
\begin{equation}
	\label{eqn:int_T_drift}
	\left|\int_t^{t+T_{\rm drift}}dt' \dot{f}_{\rm gw}(t')\right| \leq \Delta f,
\end{equation}
for $0\leq t \leq T_{\rm obs}$, where $\dot{f}_{\rm gw}$ is the first time derivative of $f_{\rm gw}$. We aim to search for signals with spin-down time-scale in the range $10^2\,{\rm s} \lesssim \tau \lesssim 10^4$\,s and $f_{\rm gw} \lesssim 2$\,kHz, such that $\dot{f}_{\rm gw}$ satisfies \mbox{$|\dot{f}_{\rm gw}| \approx f_{\rm gw}/\tau \lesssim 1$\,Hz\,s$^{-1}$} in most of the parameter space. Given \mbox{$T_{\rm drift} = 1$\,s} and a frequency bin width of \mbox{$\Delta f = 1$\,Hz}, Eqn.~(\ref{eqn:int_T_drift}) is satisfied when $|\dot{f}_{\rm gw}| \leq 1$\,Hz\,s$^{-1}$. In the full frequency band $B$ analyzed, we have $N_Q = N_O = B/\Delta f$ and $N_T = T_{\rm obs}/T_{\rm drift}$.

The HMM emission probability at each discrete time, defined as the likelihood of hidden state $q_i$ being observed in state $o_j$, is given by \cite{Suvorova2016}
\begin{equation}
	L_{o_j q_i} = P [o(t_n)=o_j|q(t_n)=q_i].
\end{equation}
The post-merger signal has a much shorter spin-down time-scale $\tau$ than a normal continuous-wave signal described in Ref. \cite{Sun2018}. The motion of the Earth with respect to the solar system barycenter (SSB) can be neglected during the interval [$t,t+T_{\rm drift}$], unlike in searches based on the $\mathcal{F}$-statistic \cite{Jaranowski1998}. Hence the HMM emission probability $L_{o(t)q_i} = P [o(t)|f_i \leq \fgw(t) \leq f_i+\Delta f]\propto \exp[G(f_i)]$ is calculated from the running-mean normalized power in short Fourier transforms (SFTs) with length $T_{\rm SFT} = T_{\rm drift} =1$\,s. We normalize the SFT power in each frequency bin by the average power in a window with width $3\Delta f = 3$\,Hz centered on the bin to reduce the impact caused by variation of the power spectrum density (PSD) in a wide frequency band 100\,Hz--2\,kHz. We write
\begin{equation}
G(f_i) = \sum_{X}\frac{3\tilde{x}^X_i \tilde{x}^{X*}_i} {\tilde{x}^X_{i-1}\tilde{x}^{X*}_{i-1} + \tilde{x}^X_i \tilde{x}^{X*}_i + \tilde{x}^X_{i+1}\tilde{x}^{X*}_{i+1}},
\end{equation}
where $i$ indexes the SFT frequency bin, $X$ indexes the detector, and the repeated index $i$ on the right-hand side does not imply summation. 

The transition probability from time $t_n$ to $t_{n+1}$ is defined as \cite{Suvorova2016}
\begin{equation}
	\label{eqn:A_matrix}
	A_{q_j q_i} = P [q(t_{n+1})=q_j|q(t_n)=q_i],
\end{equation}
which depends on the signal evolution characteristics. 
We assume that the signal frequency is monotonously decreasing and the auto-correlation time-scale of timing noise is much longer than $T_{\rm drift}$, and hence adopt the HMM transition probabilities
\begin{equation}
	\label{eqn:trans_matrix-pmr}
	A_{q_{i-1} q_i} = A_{q_i q_i} = \frac{1}{2},
\end{equation}
with all other $A_{q_j q_i}$ being zero. These choices of $A$ also imply that the signal frequency is approximated by a negatively biased random walk, consistent with a potential rapidly spin-down post-merger remnant (cf. the unbiased random walk in searches for low-mass X-ray binaries where the frequency drift is dominated by spin wandering \cite{Suvorova2016, ScoX1ViterbiO1}).
Since we have no independent knowledge of $\fgw$, we choose a uniform prior, viz.
\begin{equation}
	\Pi_{q_i} = N_Q^{-1}.
\end{equation}

The probability that the hidden state path $Q=\{q(t_0), \cdots, q(t_{N_T})\}$ gives rise to the observed sequence $O=\{o(t_0), \cdots, o(t_{N_T})\}$ via a Markov chain equals
\begin{equation}
	\label{eqn:prob}
	\begin{split}
		P(Q|O) = & L_{o(t_{N_T})q(t_{N_T})} A_{q(t_{N_T})q(t_{N_T-1})} \cdots L_{o(t_1)q(t_1)} \\ 
		& \times A_{q(t_1)q(t_0)} \Pi_{q(t_0)}.
	\end{split}
\end{equation}
The most probable path, maximizing $P(Q|O)$, is denoted by
\begin{equation}
	Q^*(O)= \arg\max P(Q|O),
\end{equation}
where $\arg \max (\cdots)$ returns the argument that maximizes the function $(\cdots)$. Here $Q^*(O)$ gives the best estimate of $q(t)$ over the total observation $T_{\rm obs} = N_T T_{\rm drift}$.

\subsection{Gravitational-wave signal model}
\label{sec:model}
The HMM described in Sec.~\ref{sec:formulation} is used as a model-agnostic search strategy for post-merger signals. However, we need a specific signal model in order to validate the method. In this subsection, we describe the signal model used to conduct simulations and derive constraints on the source properties for the rest of the paper. We emphasize that the search itself does not rely on any particular signal model.

We assume that the remnant spins down rapidly with \cite{magnetar}
\begin{equation}
\label{eqn:magnetar_freq}
\fgw(t) = \fgwini \left(1+\frac{t}{\tau}\right)^{\frac{1}{1-n}},
\end{equation}
where $\fgwini$ is the initial signal frequency at $t=0$, the coalescence time after the inspiral signal ends (e.g., the time rounded to integer GPS seconds 1187008882 for GW170817 \cite{2017-GW170817,postmerger2017}), and $n$ is the braking index defined via $\dot{f}_{\rm gw} \propto \fgw^n$. The gravitational-wave strain amplitude is given by \cite{magnetar}
\begin{equation}
\label{eqn:magnetar_waveform}
	h_0(t) = \frac{4\pi^2G}{c^4} \frac{I_{zz}\epsilon \fgwini^2}{D} \left(1+\frac{t}{\tau}\right)^{\frac{2}{1-n}},
\end{equation}
where $G$ is Newton's gravitational constant, $c$ is the speed of light, $I_{zz}$ is the principal moment of inertia, $\epsilon$ is the remnant's mass ellipticity, and $D$ is the distance to the source. The waveform (\ref{eqn:magnetar_waveform}) is consistent with signals from a magnetar, which is distorted by its strong magnetic field \cite{Cutler2002}. 

\subsection{Observing time}
\begin{figure}
	\centering
	\includegraphics[width=\columnwidth]{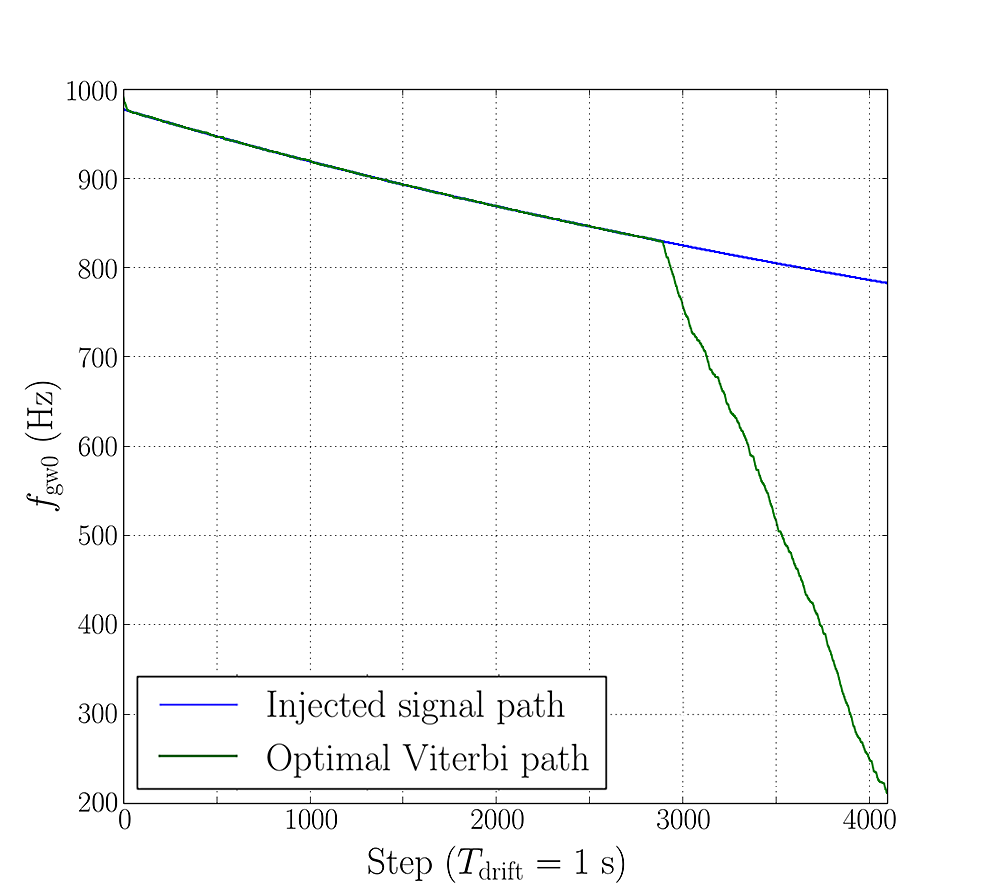}
	\caption[Injected $\fgw(t)$ and optimal Viterbi path for the injected magnetar signal.]{Injected $\fgw(t)$ (blue curve) and optimal Viterbi path (green curve) for waveform (\ref{eqn:magnetar_waveform}). Parameters: $\fgwini=1$\,kHz, $\tau = 10^4$\,s, $n=2.5$, $\epsilon = 10^{-3}$, $D=40$\,Mpc, $\cos \iota =0$, $\sqrt{S_h}=2\times 10^{-27}$\,Hz$^{-1/2}$, $T_{\rm drift}=1$\,s, and $T_{\rm obs}=4096$\,s.}
	\label{fig:pmr-sample}
\end{figure}

In most continuous-wave searches, we have $\tau \gg T_{\rm obs}$. Hence the best sensitivity is obtained, when $T_{\rm obs}$ equals the full duration of an observing run. For the post-merger remnant, however, it is better if $T_{\rm obs}$ equals the time when the signal drops below the detection limit; longer values of $T_{\rm obs}$ merely accumulate noise without improving the signal-to-noise ratio (SNR). The strain amplitude $h_0$ in (\ref{eqn:magnetar_waveform}) decreases significantly for $t \gg\tau$. Hence the SNR decreases for $T_{\rm obs} \gtrsim \tau$. Figure~\ref{fig:pmr-sample} shows an example for a magnetar signal injected into Gaussian noise, with parameters $\fgwini=1$\,kHz, $\tau = 10^4$\,s, $n=2.5$, $\epsilon = 10^{-3}$, $D=40$\,Mpc, $\cos \iota =0$, where $\iota$ is the source inclination angle, and the noise amplitude spectral density (ASD) $\sqrt{S_h}=2\times 10^{-27}$\,Hz$^{-1/2}$. We track the spin down accurately for $T_{\rm obs} \lesssim 3$\,ks and then lose it for $T_{\rm obs} \gtrsim 3$\,ks. For the last 1200\,s, the secular decrease of the estimated $\fgw$ is because of the negatively biased random walk with $A_{q_{i-1} q_i} = A_{q_i q_i} = 1/2$ in (\ref{eqn:trans_matrix-pmr}), as expected when one attempts to ``track" pure Gaussian noise. Monte-Carlo simulations show that choosing $T_{\rm obs} \sim \tau$ yields the best sensitivities for signals with $h_0$ near the detection limit. A detailed list of optimal $T_{\rm obs}$ as a function of $\tau$, $n$, and$\fgwini$ is given in Sec.~\ref{sec:sensitivity}.

\subsection{Tracking example}
\begin{figure*}
	\centering
	\subfigure[]
	{
		\label{fig:noise}
		\scalebox{0.28}{\includegraphics{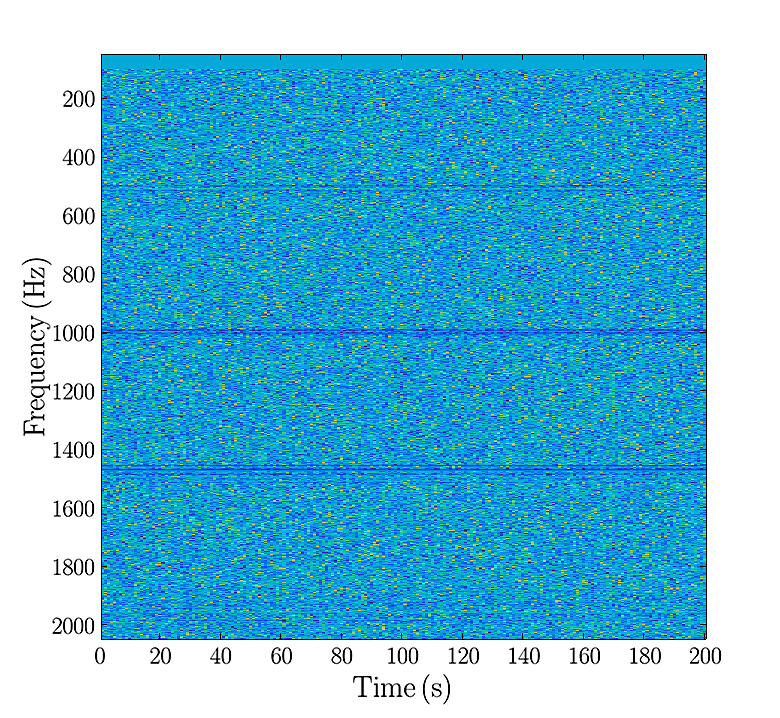}}
	}
	\subfigure[]
	{
		\label{fig:noisepath}
		\scalebox{0.36}{\includegraphics{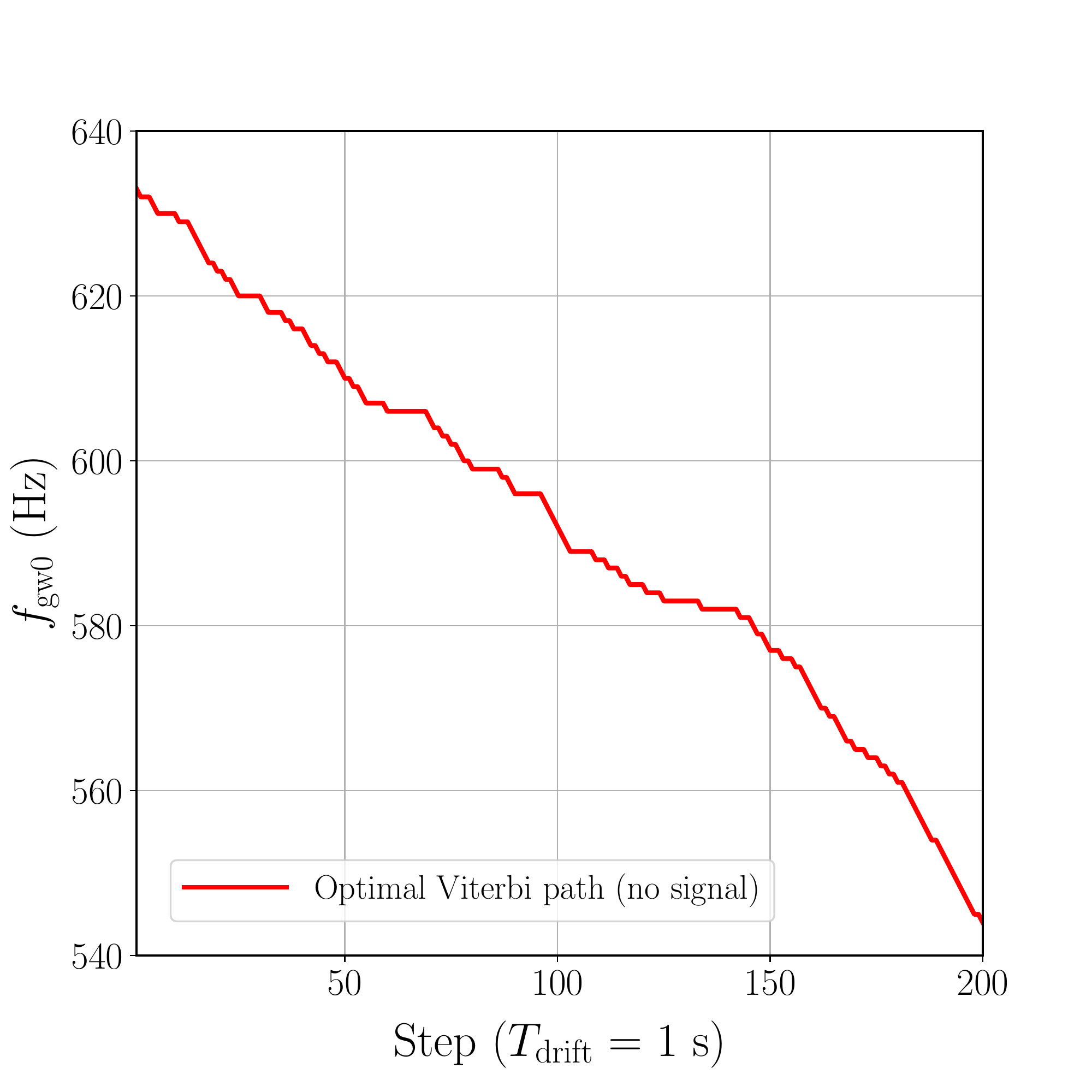}}
	}
	\subfigure[]
	{
		\label{fig:sig}
		\scalebox{0.28}{\includegraphics{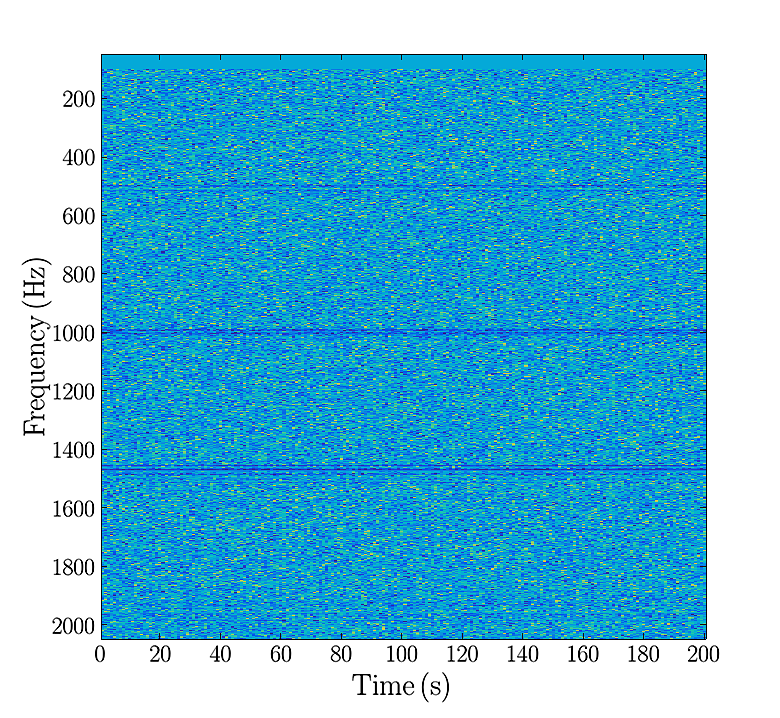}}
	}
	\subfigure[]
	{
		\label{fig:sigpath}
		\scalebox{0.36}{\includegraphics{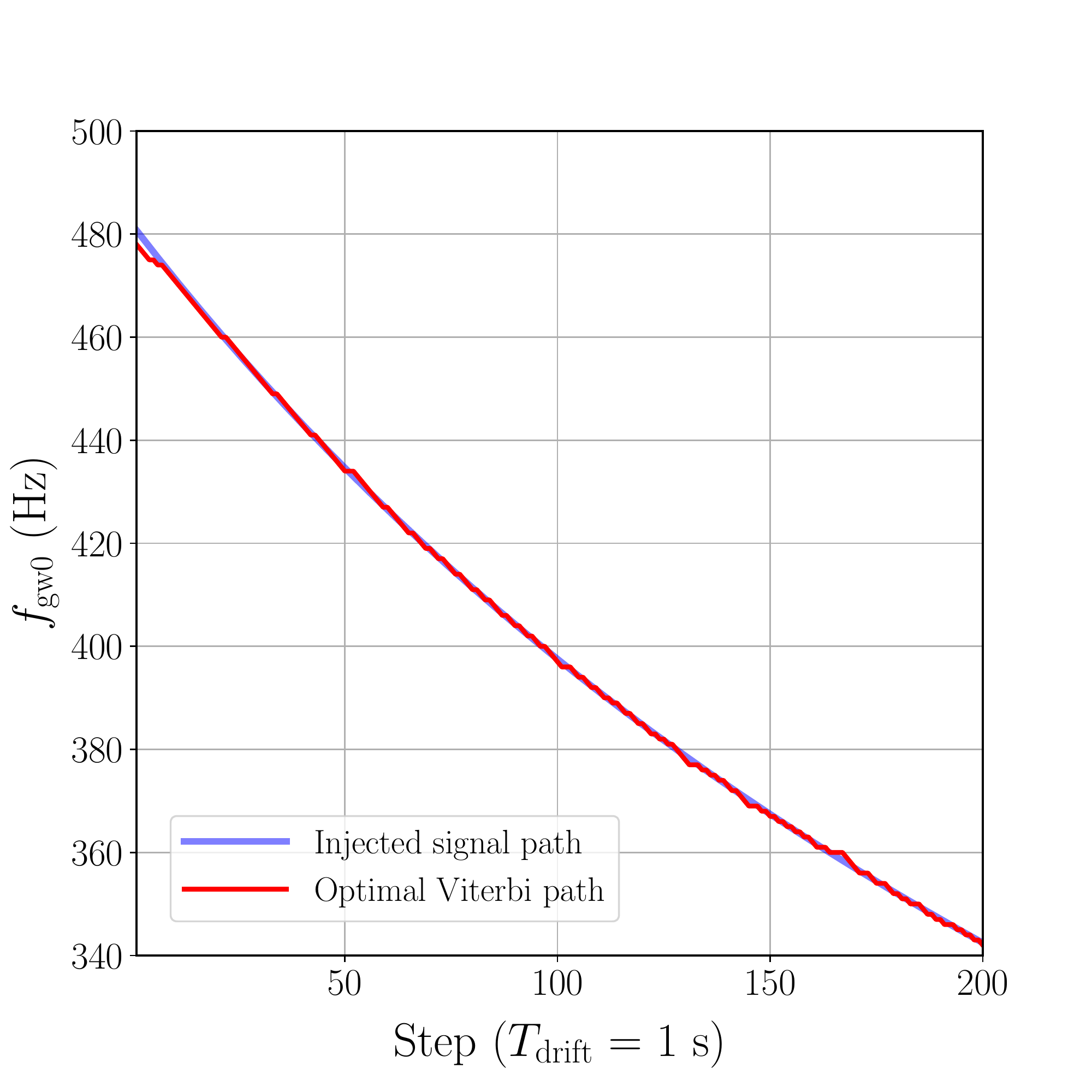}}
	}
	\caption[Spectrograms of real interferometer data without and with an injected signal, and the optimal paths returned by the tracker.]{Spectrograms of real interferometer data (a) without and (c) with an injected signal, and the optimal paths [(b) and (d)] returned by the tracker ($T_{\rm obs} =200$\,s, $N_T = 200$). The red curves in (b) and (d) represent the optimal Viterbi paths  for (a) and (c), respectively. The blue curve in (d) represents the injected signal path. A good match is obtained for an injection path hardly seen in the spectrogram. Loud instrumental lines are removed before tracking by setting $G(f_i) = 1$ if frequency bin $i$ is contaminated by lines.}
	\label{fig:real_data_sig}
\end{figure*}

Figure~\ref{fig:real_data_sig} presents a tracking example in real interferometer noise. Panels (a) and (c) show $G(f)$ spectrograms for $100\,{\rm Hz} \leq f \leq 2000$\,Hz and $0\leq t \leq 200$\,s, without and with a signal injected into instrumental noise, respectively. Loud instrumental lines are removed before tracking by setting $G(f_i) = 1$ if frequency bin $i$ is contaminated by lines, corresponding to the dark blue, horizontal strips in (a) and (c). Each 1-s-wide vertical strip is computed from one SFT at discrete times $t_0, \cdots, t_{200}$. The initial injected signal frequency is $\fgwini = 481$\,Hz in panel (c). We can hardly see any difference between the spectrums (a) and (c) because the injection is weak. The red curves in (b) and (d) represent the optimal Viterbi paths by tracking $N_T = 200$ steps using the data in (a) and (c), respectively. The detection statistic of path (b) is below threshold, consistent with a noise path; again, the frequency decreases due to the negatively biased random walk. The blue curve in (d) represents the injected signal path $\fgw(t)$. The red curve in (d) recovered by the HMM overlaps most of the blue curve, and the detection statistic lies above the threshold. We define the detection statistic, calculate the threshold, and estimate the sensitivity in Sec.~\ref{sec:implementation}.

\section{Implementation}
\label{sec:implementation}

In this section, we first define the detection statistic and calculate the threshold without referring to a chosen model (Secs.~\ref{sec:detection_statistic} and \ref{sec:threshold}). The detailed configuration and sensitivities (Secs.~\ref{sec:wait} and \ref{sec:sensitivity}) are obtained by assuming the model described in Sec.~\ref{sec:model}. 

\subsection{Detection statistic}
\label{sec:detection_statistic}
In most continuous-wave searches using HMM tracking, a detection score is defined and calculated in each 1-Hz sub-band, where the noise PSD can be regarded as flat, and the $\mathcal{F}$-statistic in $G(f)$ is normalized by the noise PSD \cite{ScoX1ViterbiO1,Sun2018}. For post-merger remnants, however, we analyze SFTs rather than the $\mathcal{F}$-statistic, and the signal can wander across a wide frequency band $\sim 10^2$\,Hz over a short observing time, in which case the variation of the noise PSD is not negligible. Hence we define a new detection statistic $\mathcal{P}$, given by 
\begin{equation}
	\mathcal{P} = \frac{1}{N_T+1}\sum_{n=0}^{N_T} G[f_{i(t_n)}],
\end{equation}
where the integer $i(t_n)$ indexes the SFT frequency bin corresponding to $q^*(t_n)$ on the optimal Viterbi path $Q^*$ ($t_0\leq t_n \leq t_{N_T}$).

\subsection{Threshold}
\label{sec:threshold}
We determine the detection threshold $\mathcal{P}_{\rm th}$ for a given false alarm probability $\alpha_{\rm f}$ and $T_{\rm obs}$ through Monte-Carlo simulations. Data sets containing pure noise are searched. For each value of $T_{\rm obs}$, we simulate 1000 noise realizations by using 1000 randomly scrambled 1-s SFT sequences from the real interferometer data, i.e., generating noise sequences by randomly permuting the SFT timestamps. The value of $\mathcal{P}$ which yields a fraction $\alpha_{\rm f}$ of positive detections is then $\mathcal{P}_{\rm th}$. We list $\mathcal{P}_{\rm th}$ ($\alpha_{\rm f}=1\%$) for $200 \leq T_{\rm obs}/(1\,{\rm s}) \leq 9688$ in the whole search band 100--2000\,Hz together with GW170817 search results in Table~\ref{tab:search_results} (Sec.~\ref{sec:results}). 

We use the randomly scrambled 1-s SFT sequences as noise realizations rather than sequential data from other observing periods mainly because the impact from the time-varying detector PSD is not negligible if the sample is taken at a time far from the event. Since the persistent instrumental lines are removed in advance, we do not expect that randomly rearranging the SFTs impacts the distribution of $\mathcal{P}$ in noise. To verify that, we draw random samples of 200-s and 600-s unscrambled sequences from six-hour data taken on the same day before the event (different data set from that analyzed in the search), obtain the detection statistics, and compare the distribution to that obtained from scrambled data. The details of the comparison are provided in Appendix~\ref{sec:noise_seq_sample}. 
\subsection{Waiting time}
\label{sec:wait}
The initial spin-down rate $|\dotfgwini|$ of a signal with $\tau \lesssim 10^3$\,s can be too high (i.e., $|\dotfgwini|>1$\,Hz\,s$^{-1}$) for Eqn.~(\ref{eqn:int_T_drift}) to be satisfied with $T_{\rm drift}=1$\,s. Table \ref{tab:wait_time} shows the $|\dotfgwini|$ values of waveforms like (\ref{eqn:magnetar_waveform}) given various $\fgwini$, $\tau$, and $n$, and the estimated waiting time $t_{\rm wait}$, after which $|\dotfgw|$ decreases such that Eqn.~(\ref{eqn:int_T_drift}) is satisfied. We start the search after waiting for a time $t_{\rm wait}$ after the merger. Alternatively, we can choose shorter $T_{\rm drift}$ (i.e., $T_{\rm drift} \leq \dotfgw^{-1/2}$) and take $t_{\rm wait} = 0$ for all waveforms. However, the sensitivity degrades because the frequency resolution $\Delta f > 1$\,Hz is relatively coarse for $T_{\rm drift} < 1$\,s. In Table \ref{tab:wait_time}, $t_{\rm wait}$ is rounded to the nearest 50\,s, because in reality we do not know the exact $t_{\rm wait}$, and Monte-Carlo simulations show that the impact on sensitivity caused by rounding is negligible. In a search without prior knowledge of the exact signal waveform, we cover the parameter space $500\,{\rm Hz} \leq \fgwini \leq 2\,{\rm kHz}$ and $2.5 \leq n \leq 7$ for $10^2\,{\rm s} \lesssim \tau \lesssim 10^4$\,s using seven discrete $t_{\rm wait}$ values in the range $0 \leq t_{\rm wait} \leq 400$\,s. The selection of $t_{\rm wait}$ is further discussed and justified in Sec.~\ref{sec:results}. 

\begin{table}
	\centering
	\setlength{\tabcolsep}{6pt}
	\renewcommand\arraystretch{1.4}
	\begin{tabular}{rrrrr}
		\hline
		\hline
		$\fgwini$ (kHz) & $\tau$ (s)& $n$ & $|\dotfgwini|$ (Hz\,s$^{-1}$)& $t_{\rm wait}$ (s)\\
		\hline
		2 & $10^2$  & 2.5 & 13.2 & 400\\
		2 & $10^2$  & 5 & 5.0 & 250\\
		2 &$10^2$ & 7 & 3.3 & 200\\
		2  & $10^3$  & 2.5 & 1.3 & 200\\
		1& $10^2$  & 2.5 & 6.6 & 200\\
		1&$10^2$  & 5 & 2.5 & 100\\
		1&$10^2$  & 7 & 1.7& 50\\
		\hline
		\hline
	\end{tabular}
	\caption[]{Initial spin-down rate $|\dotfgwini|$ and waiting time $t_{\rm wait}$ for the waveform (\ref{eqn:magnetar_waveform}) given $\fgwini$, $\tau$, and $n$.}
	\label{tab:wait_time}
\end{table}

\subsection{Sensitivity}
\label{sec:sensitivity}
Given the threshold $\mathcal{P}_{\rm th}$ determined in Sec.~\ref{sec:threshold} (see Table \ref{tab:search_results}), we evaluate the mean value\footnote{Unlike in continuous-wave searches, $h_0(t)$ of post-merger remnant signals decreases over the total observing time [see Eqn.~\eqref{eqn:magnetar_waveform}].} of the strain amplitude over the observing time, i.e., $0.5[h_0(t_{\rm wait}) + h_0(t_{\rm wait}+T_{\rm obs})]$, which yields 50\% and 90\% detection efficiencies (i.e., 50\% and 10\% false dismissal rates), denoted by $h_0^{50\%}$ and $h_0^{90\%}$, through Monte-Carlo simulations. Note that the sensitivity derived here is expressed in terms of the signal model described in Sec.~\ref{sec:model}. The injection parameters are drawn from $\tau \in \{10^2,10^3,10^4\}$\,s, $n \in \{2.5, 5, 7\}$, and $\fgwini \in \{1,2\}$\,kHz. 
For each parameter set $(\tau, n, \fgwini)$, we generate waveforms from (\ref{eqn:magnetar_waveform}) with fixed $\epsilon=0.01$ and $I_{zz}=10^{45}\,{\rm g\,cm}^2$ at distance $0.02 \leq D/{\rm Mpc} \leq 0.4$.\footnote{We choose a fixed $\epsilon=0.01$ in the simulations because generally the maximum $\epsilon$ allowed by the initial rotational energy budget is expected to be $\sim 10^{-2}$ \cite{Sarin2018}. The search sensitivity on the strain amplitude $h_0$ at the detector is not impacted by the choice of $\epsilon$. We can simply rescale among $h_0$, $\epsilon$, and $D$ using Eqn.~\eqref{eqn:magnetar_waveform}.}
The resolution is $\Delta D/{\rm Mpc} = 0.001$, 0.005, and 0.05 for ranges \mbox{$0.02 \leq D/{\rm Mpc} \leq 0.05$}, \mbox{$0.05 < D/{\rm Mpc} \leq 0.2$}, and \mbox{$0.2 < D/{\rm Mpc} \leq 0.4$}, respectively.
We run 200 injections for each $(\tau, n, \fgwini, D)$, setting random initial phase, inclination and polarization angles, and the sky location of NGC4993 (right ascension 13.1634\,hrs, declination $−23.3815^\circ$), where GW170817 is located.\footnote{Although constraints exist on the orientation of the pre-merger binary system \cite{2017-GW170817}, and the remnant's spin should be closely aligned with the orbital angular momentum of the pre-merger system, we adopt the flat prior $\cos\iota \in [-1,1]$ in the simulations, because in practice, we are unlikely to be sensitive to GW170817. We aim to validate the method for general targets with as few assumptions as possible. We defer using constrained priors on parameters like $\cos\iota$ to future events.} 
The synthetic signals are injected into scrambled interferometer noise SFT sequences (randomly permuting the SFT timestamps) in the search band 100--2000\,Hz. We run the HMM with $T_{\rm drift} =1$\,s, $t_{\rm wait}$ drawn from Table~\ref{tab:wait_time}, and various observing times $200 \leq T_{\rm obs}/(1\,{\rm s}) \leq 9688$. The optimal $T_{\rm obs}$ yielding the best sensitivity is listed in Table~\ref{tab:Tobs} for each waveform. 

\begin{figure*}
	\centering
	\subfigure[]
	{
		\includegraphics[width=\columnwidth]{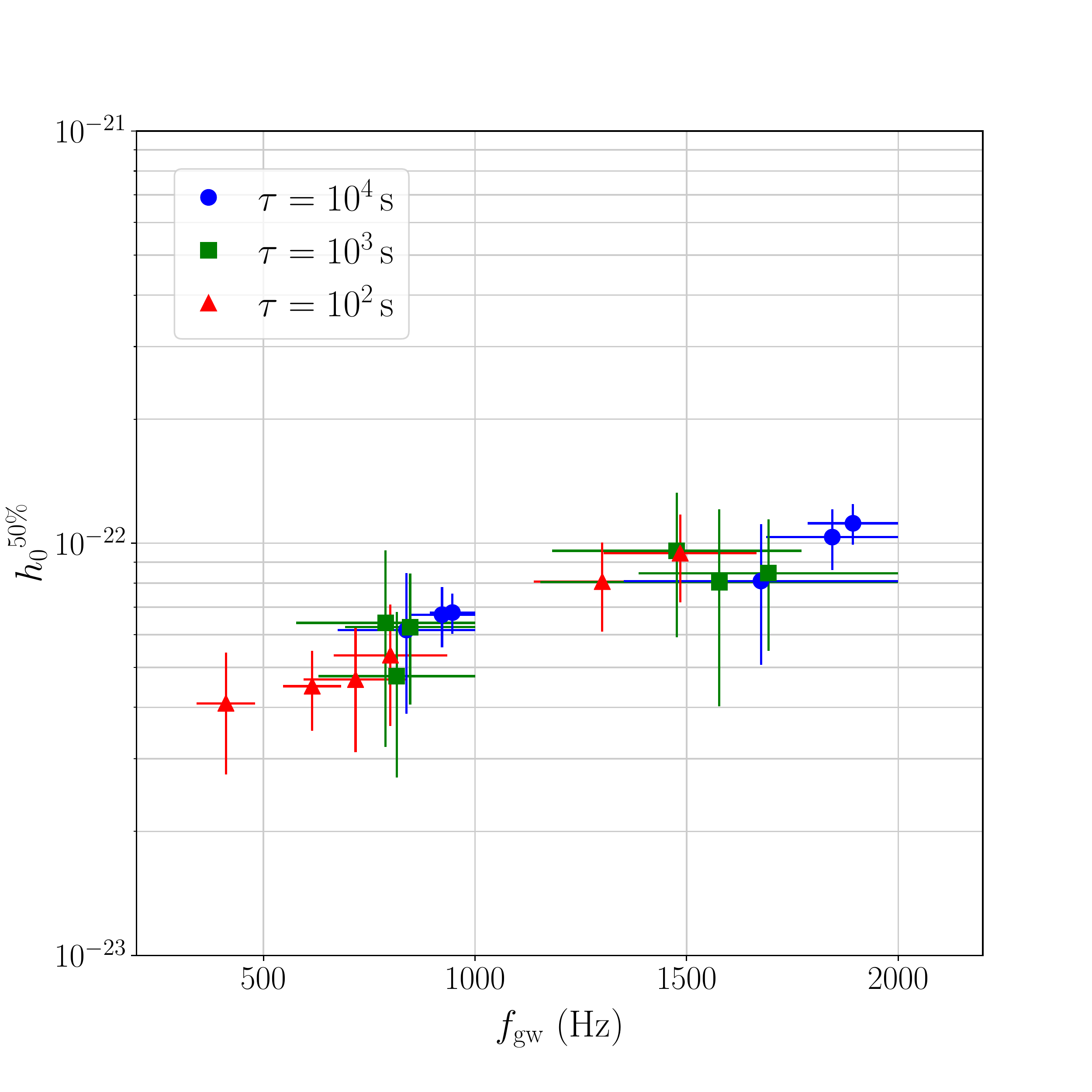}
	}
	\subfigure[]
	{
		\includegraphics[width=\columnwidth]{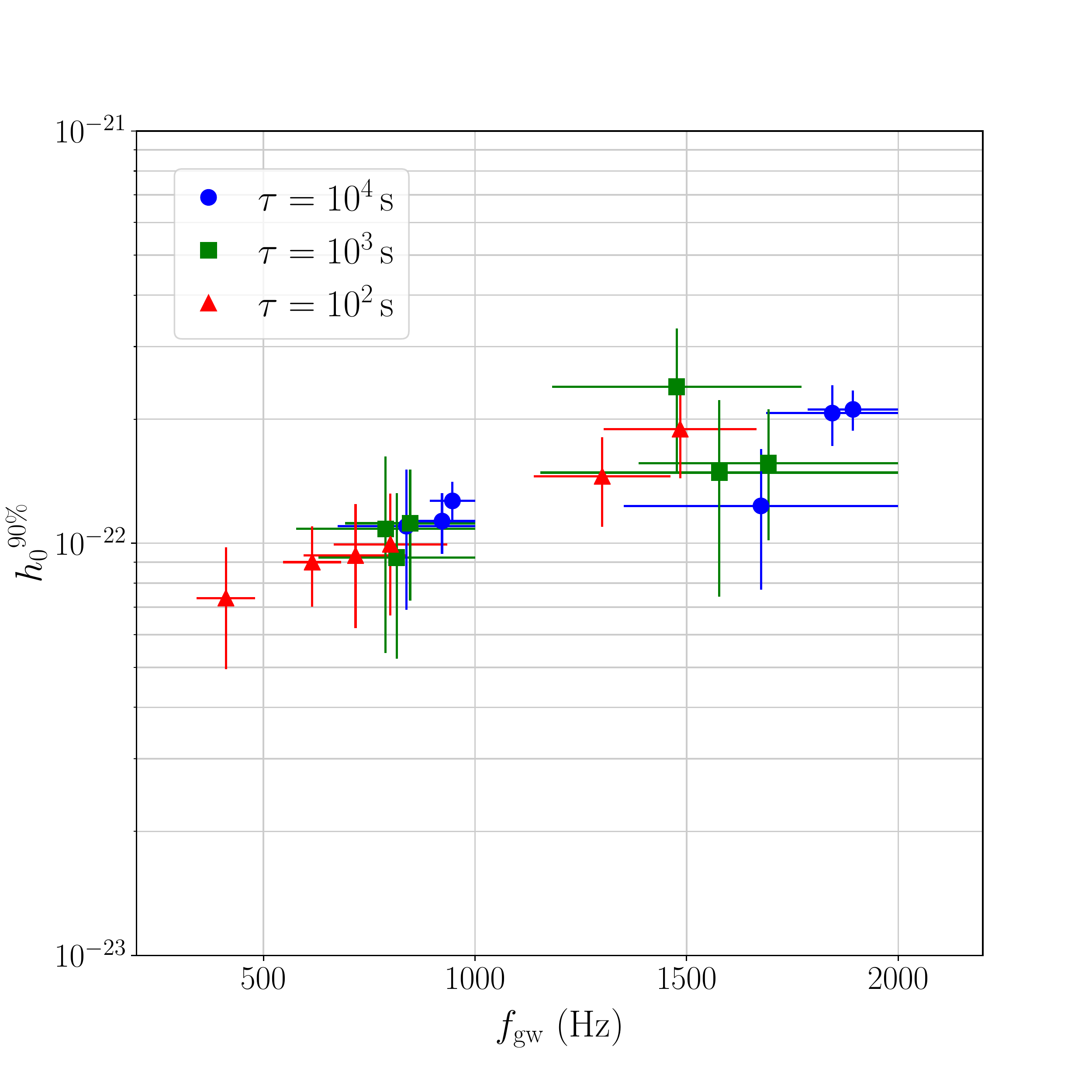}
	}
	\caption[]{Wave strain upper limits (a) $h_0^{50\%}$ and (b) $h_0^{90\%}$ as functions of injected $\fgw$ ($\alpha_{\rm f} = 1\%$). Each marker represents the upper limit for each parameter set $[\tau, n, \fgw(0)]$. Blue dots, green squares, and red triangles represent $\tau = 10^4$\,s, $10^3$\,s, and $10^2$\,s, respectively. The horizontal and vertical bars indicate the ranges spanned by $\fgw(t)$ and $h_0(t)$, as they decrease during the interval $[t_{\rm wait}, t_{\rm wait}+T_{\rm obs}]$, with the markers centered in the ranges.}
	\label{fig:h0_UL}
\end{figure*}

Figure~\ref{fig:h0_UL} displays (a) $h_0^{50\%}$ and (b) $h_0^{90\%}$ as a function of $\fgw$ for each waveform given $(\tau, n, \fgwini)$ and the optimal $T_{\rm obs}$. Blue dots, green squares, and red triangles represent $\tau = 10^4$\,s, $10^3$\,s, and $10^2$\,s, respectively. The horizontal and vertical bars correspond to the ranges covered by $\fgw(t)$ and $h_0(t)$ during the interval $[t_{\rm wait}, t_{\rm wait}+T_{\rm obs}]$, respectively.
For the injected signals described above with unknown inclination $\cos\iota$ at the sky location of GW170817, the best results we obtain are $h_0^{50\%}=4.1\times 10^{-23}$ and $h_0^{90\%}=7.4\times 10^{-23}$. In continuous-wave searches, the integration time is always long enough to average out the sensitivity variation to sources at different sky locations because of the Earth's rotation. In post-merger remnant searches, the orientation of the source with respect to the detector's antenna pattern impacts the SNR significantly for signals with duration less than a day. Hence the sensitivity needs to be recalculated for a source at a different sky location through analytic calculations or Monte-Carlo simulations.

\begin{table}
	\centering
	\setlength{\tabcolsep}{6pt}
	\renewcommand\arraystretch{1.4}
	\begin{tabular}{rrrr}
		\hline
		\hline
		 $\tau$ (s) & $n$ & $\fgwini$ (kHz) & $T_{\rm obs}$ (s)\\
		\hline
		$10^2$ & 2.5 & 1 &200 \\
		$10^2$ & 2.5 & 2 &\\
		\hline
		$10^2$&5&1 & 600\\
		$10^2$&5& 2 &\\		
		\hline
		 $10^2$& 7& 1 & 1000\\
		 $10^2$& 7& 2 &\\		
		 $10^3$& 2.5&1 &\\
		 $10^3$& 2.5& 2 &\\
		\hline
		$10^3$& 5& 1 & 8000\\
	    $10^3$& 5& 2 &\\		
		$10^3$& 7& 1 &\\
		$10^3$& 7&2 &\\
		$10^4$& 2.5& 1 &\\
		$10^4$& 2.5& 2 & \\
		\hline
	    $10^4$& 5& 1 & 9688\\
		$10^4$& 5& 2  &\\		
		$10^4$& 7 &1 &\\
		$10^4$& 7& 2 &\\
		\hline
		\hline
	\end{tabular}
	\caption[]{Optimal $T_{\rm obs}$ providing the best sensitivity for waveforms (\ref{eqn:magnetar_waveform}) given the parameters $(\tau, n, \fgwini)$ obtained from Monte-Carlo simulations.}
	\label{tab:Tobs}
\end{table}

\section{GW170817 analysis}
\label{sec:gw170817}
For the convenience of the reader, we reproduce the HMM results from Ref.~\cite{long-duration-pmr} in this paper, and describe the search configuration, procedure, and follow-up step by step.

\subsection{Data}
\label{sec:data_set}
In the search presented in Ref.~\cite{long-duration-pmr}, we analyze 9688 seconds of data from Advanced LIGO O2 run after the merger of GW170817 (GPS time 1187008882 to 1187018570) with $t_{\rm wait}$ and $T_{\rm obs}$ optimized in Tables~\ref{tab:wait_time} and \ref{tab:Tobs} for different $\tau$, $n$, and $\fgwini$. We do not analyze a longer time series because (1) several intervals in the data after GPS time 1187018570 are not in analyzable science mode (i.e., either the detectors are not in observing mode or the data quality is not good enough for conducting a search), and (2) signals with $10^2\,{\rm s} \lesssim \tau \lesssim 10^4$\,s drop below the sensitivity limit after $\sim 10^4$\,s; observing longer merely accumulates noise without improving the SNR. We generate 9688 SFTs ($T_{\rm SFT} =1$\,s) with Hann windowing \cite{MendellPresentation, Goetz2011} and input them into the search. Frequency bin $i$ is cleaned by setting $G(f_i) = 1$ before the search if loud, persistent instrumental lines lie in bin $i$. A full list of cleaned SFT bins is given in Table~\ref{tab:lines}.

\begin{table}
	\centering \footnotesize
	\setlength{\tabcolsep}{3pt}
	\renewcommand\arraystretch{1.06}
	\begin{tabular}{lll}
		\hline
		\hline
		Detector & Bin (Hz) & Line origin \\
		\hline
		H1 & 300 & Beam splitter violin mode 1st harmonic \\
		 & 302--303 & Beam splitter violin mode 1st harmonic\\
		 & 503 & Violin mode 1st harmonic\\
		 & 992 & Violin mode 2nd harmonic\\
		 & 995--996 & Violin mode 2nd harmonic\\
		 & 1006 & Violin mode 2nd harmonic\\
		 & 1009 & Violin mode 2nd harmonic\\
		 & 1456 & Violin mode 3rd harmonic\\
		 & 1562 & Violin mode 3rd harmonic\\
		 & 1468 & Violin mode 3rd harmonic \\ 
		 & 1472 & Violin mode 3rd harmonic \\
		 & 1475 & Violin mode 3rd harmonic\\
		 & 1478 & Violin mode 3rd harmonic\\
		 & 1484--1485 & Violin mode 3rd harmonic \\
		 & 1923 & Violin mode 4th harmonic \\
		 & 1932& Violin mode 4th harmonic \\
		 & 1957 & Violin mode 4th harmonic\\
		\hline
		L1 & 306 & Beam splitter violin mode 1st harmonic\\
		 & 315 & Violin mode 1st harmonic\\ 
		 & 449--450 & Violin mode 1st harmonic\\
		 & 508 & Violin mode 1st harmonic \\
		 & 511--512 & Violin mode 1st harmonic \\
		 & 517 & Violin mode 1st harmonic \\
		 & 1022 & Violin mode 2nd harmonic\\
		 & 1457--1458 & Violin mode 3rd harmonic \\
		 & 1471 & Violin mode 3rd harmonic\\
		 & 1492 & Violin mode 3rd harmonic \\
		 & 1496 & Violin mode 3rd harmonic\\
		 & 1499 & Violin mode 3rd harmonic\\
		 & 1505--1506 & Violin mode 3rd harmonic\\
		 & 1511 & Violin mode 3rd harmonic \\
		 & 1922 & Violin mode 4th harmonic \\
		 & 1941 & Violin mode 4th harmonic \\
		 & 1958 & Violin mode 4th harmonic \\
		 & 1962 & Violin mode 4th harmonic\\
		 & 1967 & Violin mode 4th harmonic\\
		 & 1973 & Violin mode 4th harmonic \\
		 & 1982--1983 & Violin mode 4th harmonic \\
		 & 1986 & Violin mode 4th harmonic \\
		\hline
		\hline
	\end{tabular}
	\caption[]{SFT bins cleaned before the search and the origin of the instrumental line contamination.}
	\label{tab:lines}
\end{table}

\subsection{Results}
\label{sec:results}
We search the data described in Sec.~\ref{sec:data_set} in the frequency band 100--2000\,Hz. As we have no prior knowledge of signal parameters, 18 values of $T_{\rm obs}$ and seven values of $t_{\rm wait}$ are used to cover the ranges $10^2\,{\rm s} \leq \tau \leq 10^4$\,s, $2.5\leq n \leq 7$, and $500\,{\rm Hz}\leq \fgwini \leq 2$\,kHz and accommodate small deviations from waveform (\ref{eqn:magnetar_waveform}) due to uncertainties in signal models. Observing $T_{\rm obs} > 10^3$\,s yields better sensitivity for signals with $\tau > 10^3$\,s than those with $\tau < 10^3$\,s (see optimal $T_{\rm obs}$ in Table~\ref{tab:Tobs}). If a signal with $\tau < 10^3$\,s can be detected using $T_{\rm obs} > 10^3$\,s, it should also be detected with higher $\mathcal{P}$ using $T_{\rm obs} < 10^3$\,s. The initial spin-down rates of the signals calculated from Eqn.~(\ref{eqn:magnetar_freq}) for $\tau \gtrsim 10^3$\,s and $\fgwini \leq 2$\,kHz are moderate, e.g., $|\dotfgwini|<1$\,Hz\,s$^{-1}$. Hence we set $t_{\rm wait}=0$ for $T_{\rm obs} > 10^3$\,s. 

\begin{figure*}
	\centering
	\subfigure[]{\label{fig:background_200}
		\includegraphics[width=\columnwidth]{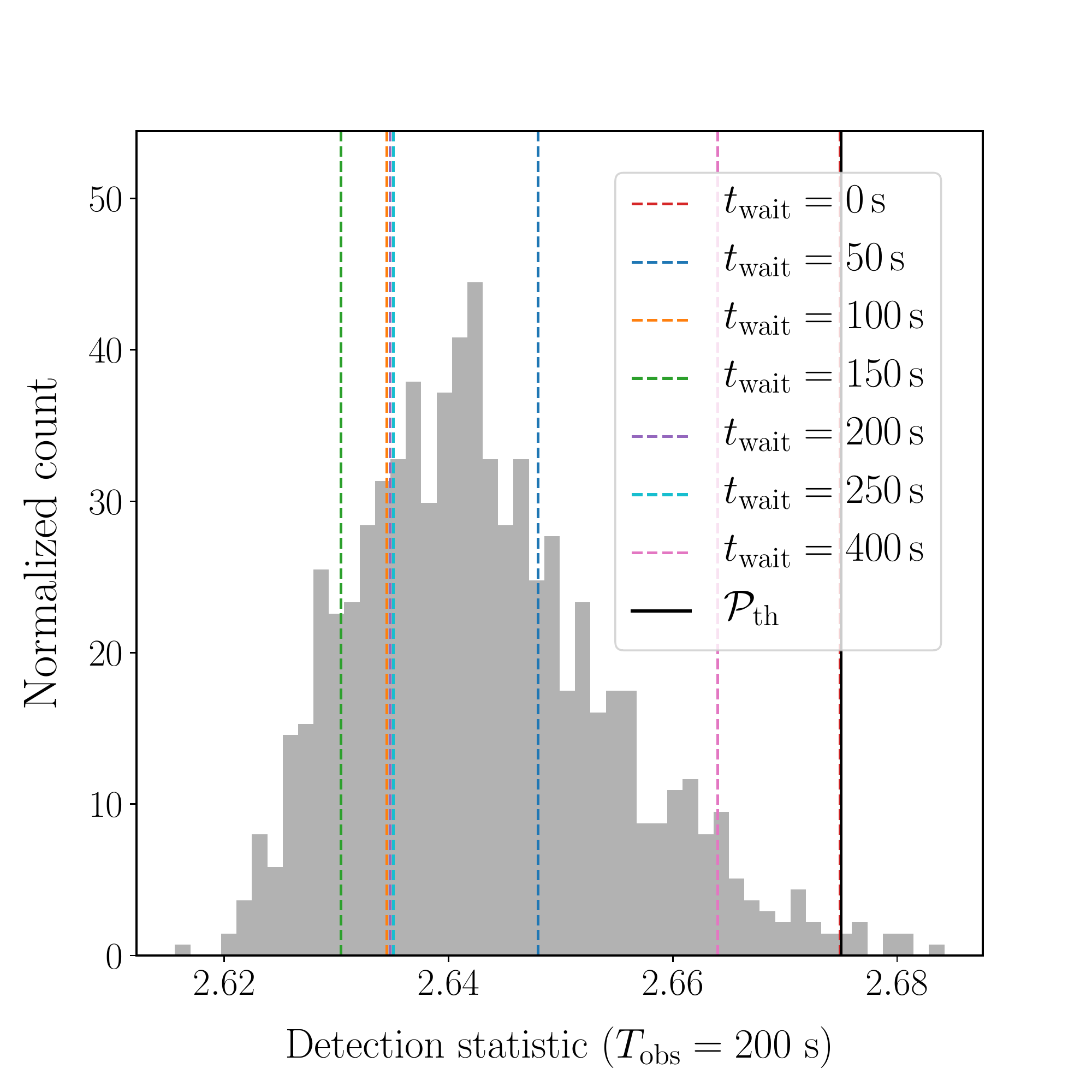}}	
	\subfigure[]{\label{fig:background_600}
		\includegraphics[width=\columnwidth]{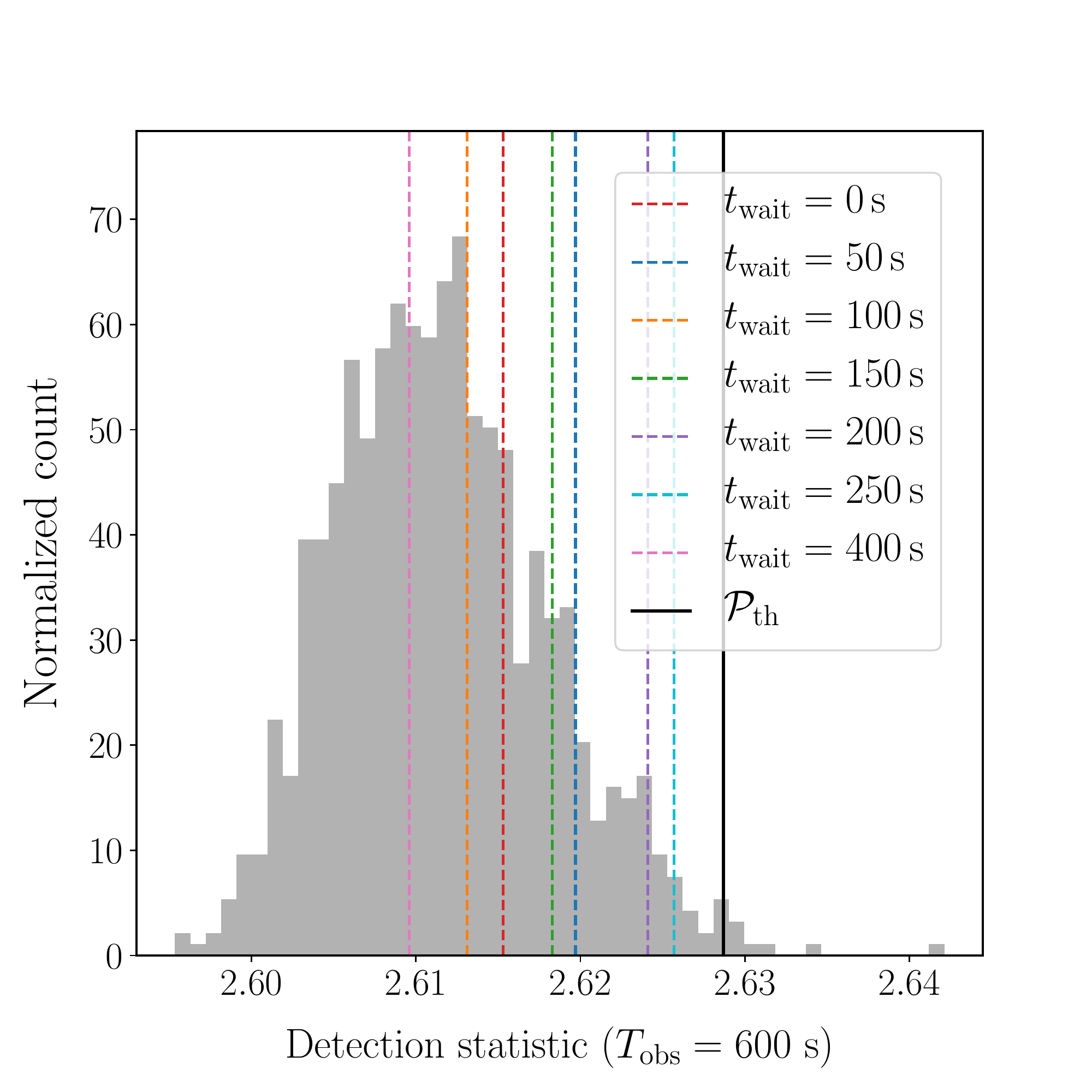}}
	\caption[]{Noise-only distribution of the detection statistics (gray histogram) and $\mathcal{P}$ values obtained from the search targeting GW170817 (colored dashed lines), using various $t_{\rm wait}$ values for (a) $T_{\rm obs} = 200$\,s and (b) $T_{\rm obs} = 600$\,s. The black solid line indicates the threshold $\mathcal{P}_{\rm th}$ ($\alpha_{\rm f} = 1\%$). The noise-only distribution is obtained from 1000 noise realizations. All $\mathcal{P}$ values are below $\mathcal{P}_{\rm th}$, consistent with the noise-only distribution. The full set of the plots corresponding to all $T_{\rm obs}$ values can be found in Appendix~\ref{sec:background}.}
	\label{fig:background_sample}
\end{figure*}

Table \ref{tab:search_results} lists all $T_{\rm obs}$ and $t_{\rm wait}$ configurations, $\mathcal{P}_{\rm th}$ for each configuration, and the resulting  $\mathcal{P}$ from the search targeting GW170817. Fig.~\ref{fig:background_sample} shows the noise-only distribution of $\mathcal{P}$ from 1000 noise realizations (gray histogram), the threshold $\mathcal{P}_{\rm th}$ corresponding to $\alpha_{\rm f} = 1\%$ (black solid line), and the resulting $\mathcal{P}$ from the GW170817 search using various $t_{\rm wait}$ values (colored dashed lines), for (a) $T_{\rm obs} = 200$\,s and (b) $T_{\rm obs} = 600$\,s. The full set of the plots corresponding to all $T_{\rm obs}$ values can be found in Appendix~\ref{sec:background}.

No candidate is found with $\mathcal{P} \geq \mathcal{P}_{\rm th}$ ($\alpha_{\rm f} = 1\%$). Only one trigger closely approaches the threshold where a candidate is deemed significant enough for further study, viz. $\mathcal{P}=2.6749$ for $T_{\rm obs} = 200$\,s and $t_{\rm wait}=0$, cf. $\mathcal{P}_{\rm th}=2.6750$ [see the rightmost red dashed line in Fig.~\ref{fig:background_200}, which almost overlaps the black solid line]. We expect that $T_{\rm obs} = 200$\,s is an optimal observing time for short-duration signals with $\tau \sim 10^2$\,s, which require $t_{\rm wait} \sim 10^2$\,s for $|\dotfgw|$ to decrease before the frequency can be correctly tracked. In other words, signals detectable with $t_{\rm wait} = 0$ satisfy Eqn.~(\ref{eqn:int_T_drift}) right after the merger. From Monte-Carlo simulations (Table \ref{tab:Tobs}), the optimal $T_{\rm obs}$ exceeds 200\,s for signals with $|\dotfgwini|<1$\,Hz\,s$^{-1}$. Hence if a signal yields $\mathcal{P}$ close to $\mathcal{P}_{\rm th}$ for $T_{\rm obs} = 200$\,s and $t_{\rm wait}=0$, we should obtain $\mathcal{P} \gtrsim \mathcal{P}_{\rm th}$ using longer $T_{\rm obs} > 200$\,s and $t_{\rm wait}=0$. But no $\mathcal{P}$ value for $300\,{\rm s} \leq T_{\rm obs} \leq 1000$\,s with $t_{\rm wait}=0$ is close to $\mathcal{P}_{\rm th}$ (see Fig.~\ref{fig:background2} in Appendix.~\ref{sec:background}). Hence we do not expect $\mathcal{P}=2.6749$ ($T_{\rm obs} = 200$\,s, $t_{\rm wait}=0$) to be of astrophysical origin. 

The above follow-up refers to the simulation results using the model described in Sec.~\ref{sec:model}. The argument, however, is generally valid for signals from a rapidly spin-down remnant. The HMM strategy is designed to accommodate uncertainties in the signal model. The threshold does not depend on a specific choice of signal model. It is likely that the relatively higher, subthreshold $\mathcal{P}=2.6749$ for $T_{\rm obs} = 200$\,s and $t_{\rm wait}=0$ is caused by increased instrumental noise during that particular period. In future searches, if above-threshold candidates are obtained, a veto procedure is required, e.g., examining the consistency among multiple detectors \cite{ScoX1ViterbiO1}. 

We also show in Fig.~\ref{fig:background_600} the results from $T_{\rm obs} = 600$\,s, which corresponds to the parameter space that the method is most sensitive to (e.g., $\tau=100$\,s, $n=5$). The resulting $\mathcal{P}$ values are consistent with the noise-only distribution.

\begin{table}
	\centering \footnotesize
	\setlength{\tabcolsep}{10pt}
	\renewcommand\arraystretch{1.06}
	\begin{tabular}{rlrl}
		\hline
		\hline
		$T_{\rm obs}$ (s) &  $t_{\rm wait}$ (s)& $\mathcal{P}_{\rm th}$ & $\mathcal{P}$ \\
		\hline
		200	&	0	&2.6750&	2.6749 \\
				&	50	&	&2.6480 \\
				&	100	&	&2.6345 \\
				&150	&	&2.6304 \\
				&200	&	&2.6348 \\
				&250	&	&2.6351 \\
				&400	&	&2.6640 \\
		300	&	0	&  2.6509&	2.6365 \\
				&50	&&	2.6299 \\
				&100	&&	2.6145 \\
				&150	&&	2.6208 \\
				&200	&&	2.6287 \\
				&250	&&	2.6252 \\
				&400	&&	2.6412 \\
		400		&0	& 2.6440&	2.6229 \\
				&50	&&	2.6198 \\
				&100	&&	2.6113 \\
		 		&150	&&	2.6139 \\
				&200	&&	2.6220 \\
				&250	&&	2.6255 \\
				&400	&&	2.6273 \\
		500		&0	& 2.6356&	2.6161 \\
				&50	&&	2.6120 \\
				&100	&&	2.6166 \\
				&150	&&	2.6167 \\
				&200	&&	2.6207 \\
				&250	&&	2.6260 \\
				&400	&&	2.6205 \\
		600		&0	& 2.6287&	2.6153 \\
				&50	&&	2.6197 \\
				&100	&&	2.6131 \\
				&150	&&	2.6183 \\
				&200	&&	2.6241 \\
				&250	&&	2.6257 \\
				&400	&&	2.6096 \\
		800		&0	& 2.6186&	2.6148 \\
				&50	&&	2.6159 \\
				&100	&&	2.6154 \\
				&150	&&	2.6137 \\
				&200	&&	2.5989 \\
				&250	&&	2.5932 \\
				&400	&&	2.6108 \\
		1000	&	0	&2.6088&	2.5960 \\
				&50	&&	2.5918 \\
				&100	&&	2.5924 \\
				&150	&&	2.5918 \\
				&200	&&	2.5974 \\
				&250	&&	2.5958 \\
				&400	&&	2.5964 \\
		1500	&	0	&2.6025&	2.5948 \\
		2000	&	0	&2.5962&	2.5945 \\
		2500	&	0	&2.5919&	2.5894 \\
		3000	&	0	&2.5869&	2.5788 \\
		4000	&	0	&2.5782&	2.5777 \\
		5000	&	0	&2.5615&	2.5549 \\
		6000	&	0	&2.5380&	2.5344 \\
		7000	&	0	&2.5131&	2.5074 \\
		8000	&	0	&2.4894&	2.4831 \\
		9000	&	0	&2.4687&	2.4632 \\
		9688	&	0	&2.4556&	2.4499 \\
		\hline
		\hline
	\end{tabular}
	\caption[]{Detection statistic $\mathcal{P}$ (last column) for HMM searches of Advanced LIGO data collected after the merger of GW170817 (GPS time 1187008882 to 1187018570) using different $T_{\rm obs}$ and $t_{\rm wait}$ (search frequency band: 100--2000\,Hz). Detection threshold $\mathcal{P}_{\rm th}$ for false alarm probability $\alpha_{\rm f}=1\%$ is listed in Column 3. No trigger is found above $\mathcal{P}_{\rm th}$.}
	\label{tab:search_results}
\end{table}

\subsection{Upper limits}
\label{sec:UL}

We now convert $h_0^{90\%}$ from Sec.~\ref{sec:sensitivity} into astrophysical upper limits. The full results are presented and discussed in Ref.~\cite{long-duration-pmr}. We focus on a scenario where the spin down of the remnant is dominated by gravitational-wave emission from a static quadrupole deformation, i.e., $n=5$ in Eqns.~(\ref{eqn:magnetar_freq}) and (\ref{eqn:magnetar_waveform}). 
Note that we cover a finer grid of \mbox{$\fgwini \in \{500, 750, 1000, 1250, 1500, 1750, 2000\}$\,Hz} for $\tau \in \{10^2, 10^3, 10^4\}$\,s and fixed $n=5$ in the same way as described in Sec.~\ref{sec:sensitivity}, motivated by the parameter space covered in Ref.~\cite{long-duration-pmr}. We record $h_0^{90\%}$ for each $(\tau, \fgwini)$.

\begin{figure}
	\centering
	\includegraphics[width=\columnwidth]{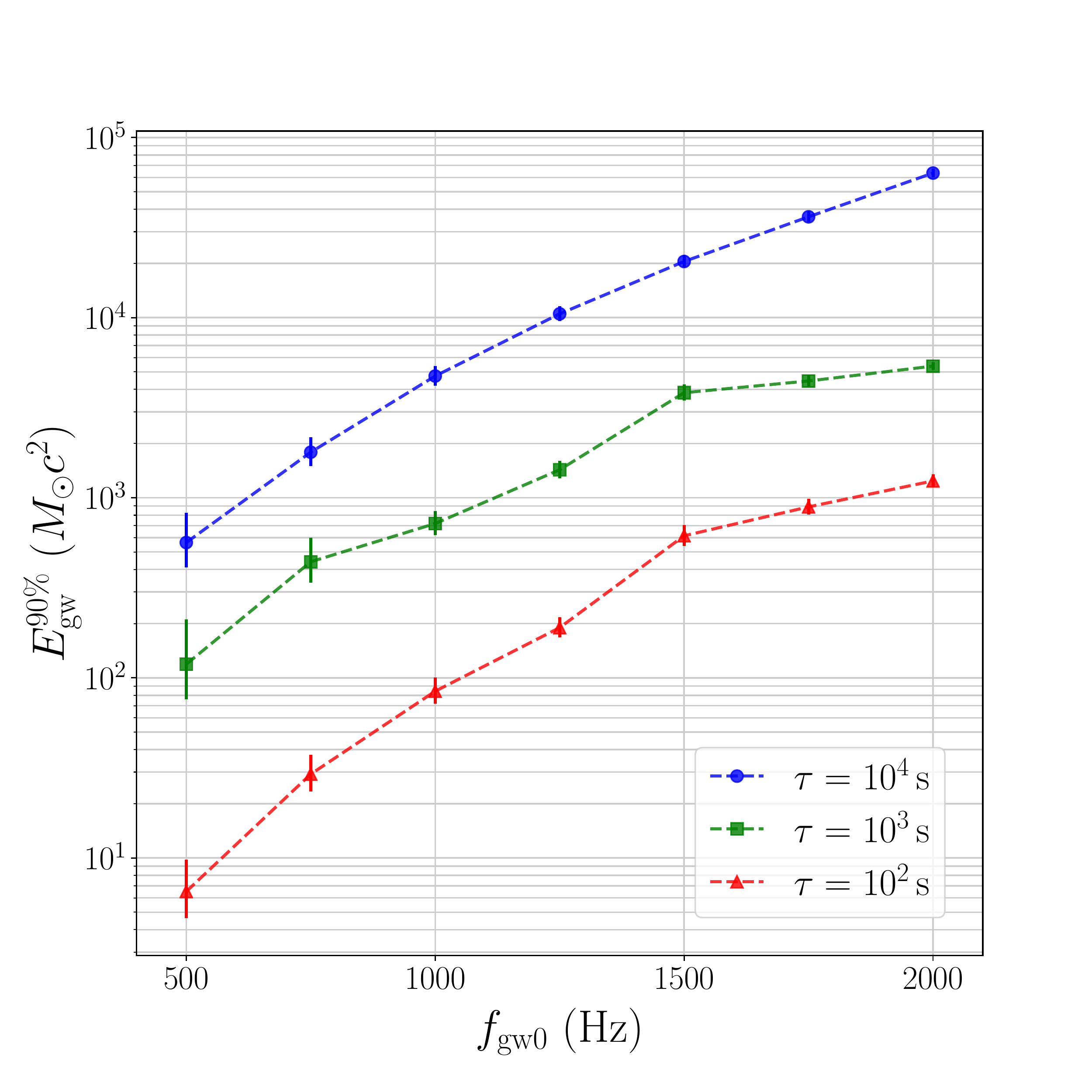}
	\caption[]{Upper limits on energy at 90\% confidence, $E_{\rm gw}^{90\%}$, as a function of $\fgwini$, assuming $I_{zz}=4.337\times10^{45}\,{\rm g\,cm}^2$ and unknown orientation. Blue dots, green squares, and red triangles represent the results obtained in this search for signals with $\tau = 10^4$\,s, $10^3$\,s, and $10^2$\,s, respectively. Vertical error bars indicate uncertainties. Some of the error bars are too small to be seen by eye. This figure is equivalent to Fig.~7 in Ref.~\cite{long-duration-pmr}.}
	\label{fig:egw_UL}
\end{figure}

The total energy radiated in the form of gravitational waves up to time $t$ is given by \cite{Sarin2018}
\begin{equation}
\label{eqn:energy}
E_{\rm gw}(t) =  \frac{32 \pi^6 G I_{\rm zz}^2 f_{\rm gw0}^6 \epsilon^2 \tau}{5c^5} \frac{n-1}{n-7} \left[1-\left(1+\frac{t}{\tau}\right)^{\frac{7-n}{1-n}} \right].
\end{equation}
Without a detection, we can derive the 90\% confidence upper limit on $E_{\rm gw}(t \to \infty)$, denoted by $E_{\rm gw}^{90\%}$, given $h_0^{90\%}$ and the distance to GW170817 ($40^{+8}_{-14}$\,Mpc) \cite{2017-GW170817}. 
Fig.~\ref{fig:egw_UL} displays $E_{\rm gw}^{90\%}$ as a function of $\fgwini$. Blue dots, green squares, and red triangles correspond to signals with $\tau = 10^4$\,s, $10^3$\,s, and $10^2$\,s, respectively. The vertical error bars indicate the uncertainty. Some of the error bars are too small to be seen by eye. Fig.~\ref{fig:egw_UL} is equivalent to Fig.~7 in Appendix A of Ref.~\cite{long-duration-pmr}. The lowest upper limit we obtain is $E_{\rm gw}^{90\%} = 6.51 M_\odot c^2$ for $\fgwini=500$\,Hz and $\tau=10^2$\,s, assuming unknown source orientation. Given the system mass is $2.73^{+0.04}_{-0.01} M_\odot$ \cite{Abbott:2018wiz}, the constraint of $E_{\rm gw}^{90\%}$ obtained from the search exceeds the total rest energy of the system and hence is uninformative.

\subsection{Astrophysical reach of the search}
\label{sec:distance}

\begin{figure}
	\centering
	\includegraphics[width=\columnwidth]{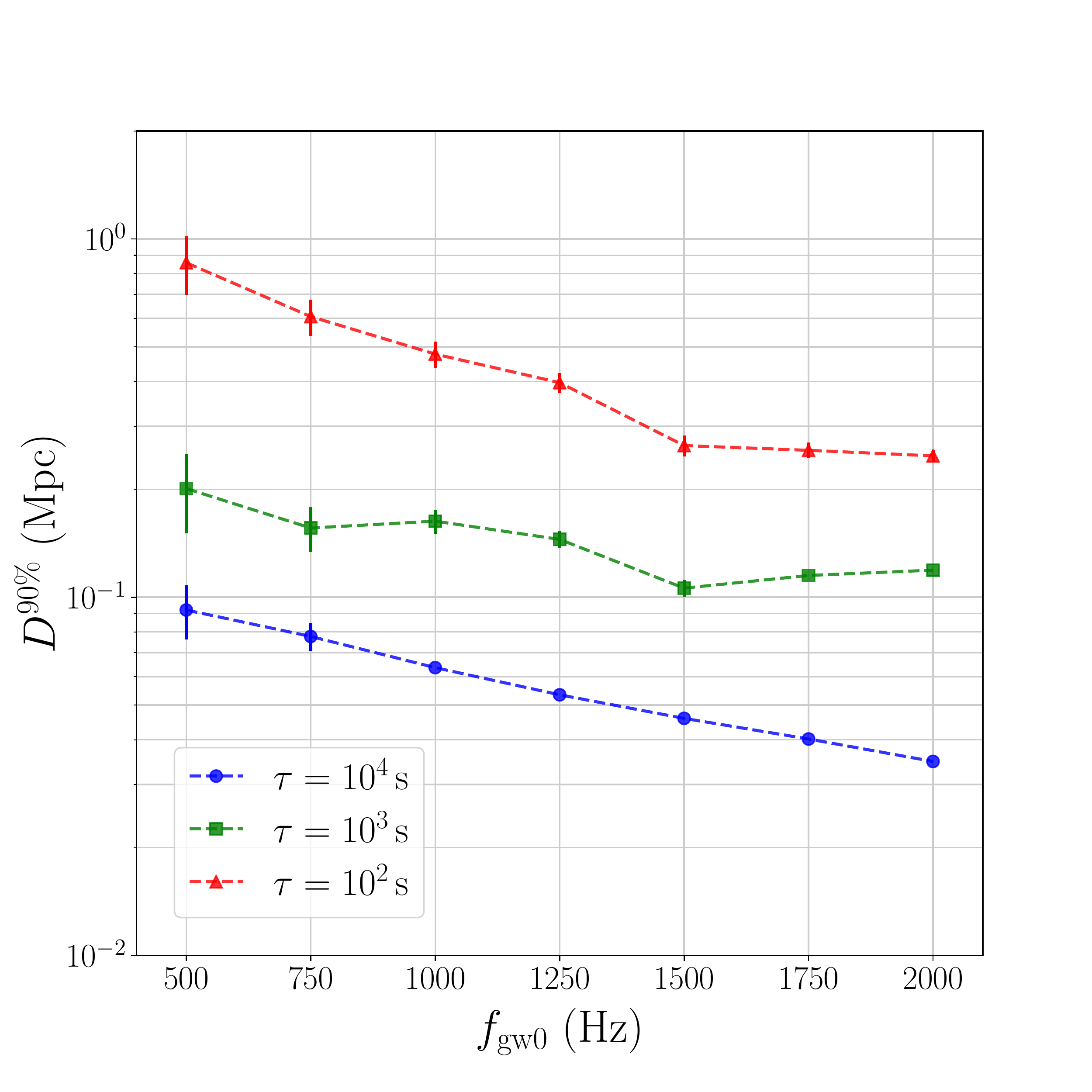}
	\caption[]{Astrophysical reach of the search at 90\% confidence, $D^{90\%}$, as a function of $\fgwini$ for maximum allowed ellipticity $\epsilon_{\rm max}$, unknown orientation, and $I_{zz}=4.337\times10^{45}\,{\rm g\,cm}^2$. Blue dots, green squares, and red triangles represent the results obtained in this search for signals with $\tau = 10^4$\,s, $10^3$\,s, and $10^2$\,s, respectively. Vertical error bars indicate uncertainties. Some of the error bars are too small to be seen by eye. This figure is equivalent to Fig.~7 in Ref.~\cite{long-duration-pmr}.}
	\label{fig:sensi_dd}
\end{figure}

The maximum ellipticity in Eqns.~\eqref{eqn:magnetar_waveform} and \eqref{eqn:energy}, $\epsilon_{\rm max}$, can be calculated from energy conservation, i.e., the total energy emitted in gravitational waves $E_{\rm gw}(t\to \infty)$ cannot exceed the initial rotational energy of the system $0.5 \pi^2 I_{zz} \fgwini^2 $, assuming that $\fgwini$ is twice the initial spin frequency of the remnant \cite{Sarin2018}. 
We then rescale the distance, which yields $h_0^{90\%}$, using $\epsilon_{\rm max}$ and $I_{zz}=4.337\times10^{45}\,{\rm g\,cm}^2$, a fiducial value preferred by the posteriors derived from the inspiral \cite{Abbott:2018wiz}, to obtain the astrophysical reach at 90\% confidence, denoted by $D^{90\%}$, for a hypothetical object with the maximum ellipticity at the sky location of GW170817 (Figures~\ref{fig:sensi_dd}). The largest $D^{90\%}$ we obtain is $0.86^{+0.16}_{-0.16}$\,Mpc for $\fgwini=500$\,Hz and $\tau=10^2$\,s, assuming $\epsilon_{\rm max} = 7.33\times10^{-2}$ and unknown source orientation (one to two orders of magnitude closer than GW170817).

\section{Conclusion}
\label{sec:conclusion}

In this paper, we describe how to revise the HMM tracking method used in continuous-wave searches to search for long-transient gravitational wave signals from a binary neutron star merger in order to ensure that the GW170817 search results described in Ref.~\cite{long-duration-pmr} can be fully reproduced and future analyses of post-merger remnants can be conducted by independent parties. For an event at the sky location of GW170817, the gravitational-wave strain sensitivities $h_0^{50\%}=4.1\times 10^{-23}$ and $h_0^{90\%}=7.4\times 10^{-23}$ at 50\% and 90\% confidence levels are obtained through Monte-Carlo simulations for unknown source orientation. 

No candidate is found above the detection threshold in data spanning 9688\,s after the coalescence GW170817 in Advanced LIGO O2 \cite{long-duration-pmr}. The 90\% confidence upper limit on energy radiated in gravitational waves obtained from the analysis is $E_{\rm gw}^{90\%} = 6.51 M_\odot c^2$ at the true distance of GW170817, 40\,Mpc. This corresponds to an astrophysical reach of $D^{90\%} = 0.86^{+0.16}_{-0.16}$\,Mpc for an object at the same sky location as GW170817 with maximum ellipticity $\epsilon_{\rm max} = 7.33\times10^{-2}$, $I_{zz}=4.337\times10^{45}\,{\rm g\,cm}^2$, and $n=5$. These constraints are obtained by assuming the signal model described in Sec.~\ref{sec:model}. The full results from this analysis are presented in Ref.~\cite{long-duration-pmr}, together with the results from the STAMP \cite{Thrane2011-stamp,Thrane:2013bea,Thrane:2014bma}, Adaptive Transient Hough \cite{Oliver:2018dpt,Krishnan:2004sv,T070124}, and FrequencyHough \cite{Miller:2018genfreqhough,Palomba:2005fp,Antonucci:2008jp,Astone:2014esa} analyses.

Assuming that the loss of rotational energy of the star since its birth is dominated by gravitational radiation, the indirect age-based wave strain upper limit is given by \cite{wette08}
\begin{equation}
\label{eqn:h0_age}
h_0^{\rm age}  = 2.2\times 10^{-24}\left(\frac{1\,{\rm kpc}}{D}\right)\left(\frac{1\,{\rm kyr}}{\tau}\right)^{1/2} \left(\frac{I_{zz}}{10^{45}\,{\rm g\,cm}^2}\right)^{1/2}.
\end{equation}
By substituting  $D=40$\,Mpc and $\tau = 10^4$\,s in Eqn.~(\ref{eqn:h0_age}), we obtain $h_0^{\rm age}  = 1.0\times 10^{-25}$, which is about two orders of magnitude below the sensitivity of the HMM, consistent with the search results. Although $h_0^{\rm age}$ is $\sim 10^2$ times higher for $\tau \sim 1$\,s, the spin down is too rapid for the HMM to track. 

We emphasize that the search described in this paper is not sensitive to any post-merger signal from GW170817 at 40\,Mpc in the whole parameter space studied with the current detector sensitivity.
The astrophysical reach of $\sim 1$\,Mpc is obtained in the optimal scenario with $\epsilon_{\rm max} \sim 10^{-2}$ \cite{Sarin2018}.
The search method, however, is verified and applicable to any similar events which may be observed in upcoming observing runs.
Given that the instrumental upgrades of Advanced LIGO and Virgo towards their design sensitivities \cite{ALIGO2015,Aasi-design,T1800044-design-curve} and further enhancements of LIGO A+ \cite{T1800042-Voyager} are planned to improve the strain sensitivity by a factor of 2--4, the chance of detecting post-merger signals from binary neutron star coalescences in the upcoming observing runs is still small. 
Third generation detectors (e.g., the Einstein Telescope and Cosmic Explorer \cite{Hild:2010id,Sathyaprakash:2012jk,Punturo:2010zz,Abbott2017-nextGen-CE}), however, are expected to improve the strain sensitivity by a factor of $\sim 20$--30 relative to Advanced LIGO (i.e., $\sim 40$--60 times better than O2). At that stage, observation of post-merger signals becomes promising, given the probability of detecting a couple of events like GW170817 at tens of Mpc over a few years \cite{Abbott2018catalog}.

\section{Acknowledgments}

We are grateful to Karl Wette and Grant Meadors for their comprehensive formal review of the code and validation of the method. 
We are also grateful to David Keitel, the LVC post-merger analysis group, and the Continuous Wave Working Group for detailed comments and informative discussions. 
L. Sun is a member of the LIGO Laboratory.
LIGO was constructed by the California Institute of Technology and Massachusetts Institute of Technology with funding from the National Science Foundation, and operates under cooperative agreement PHY--0757058. Advanced LIGO was built under award PHY--0823459. 
L. Sun was supported by an Australian Research Training Program Stipend Scholarship and the Albert Shimmins Fund at earlier stages of this project. 
The research is also supported by Australian Research Council (ARC) Discovery Project DP170103625 and the ARC Centre of Excellence for Gravitational Wave Discovery CE170100004. 
This paper carries LIGO Document Number LIGO--P1800291.

\appendix
\section{Noise samples from scrambled and sequential data}
\label{sec:noise_seq_sample}
\begin{figure*}
	\centering
	\subfigure[]{\label{fig:background_seq_200}
		\includegraphics[width=\columnwidth]{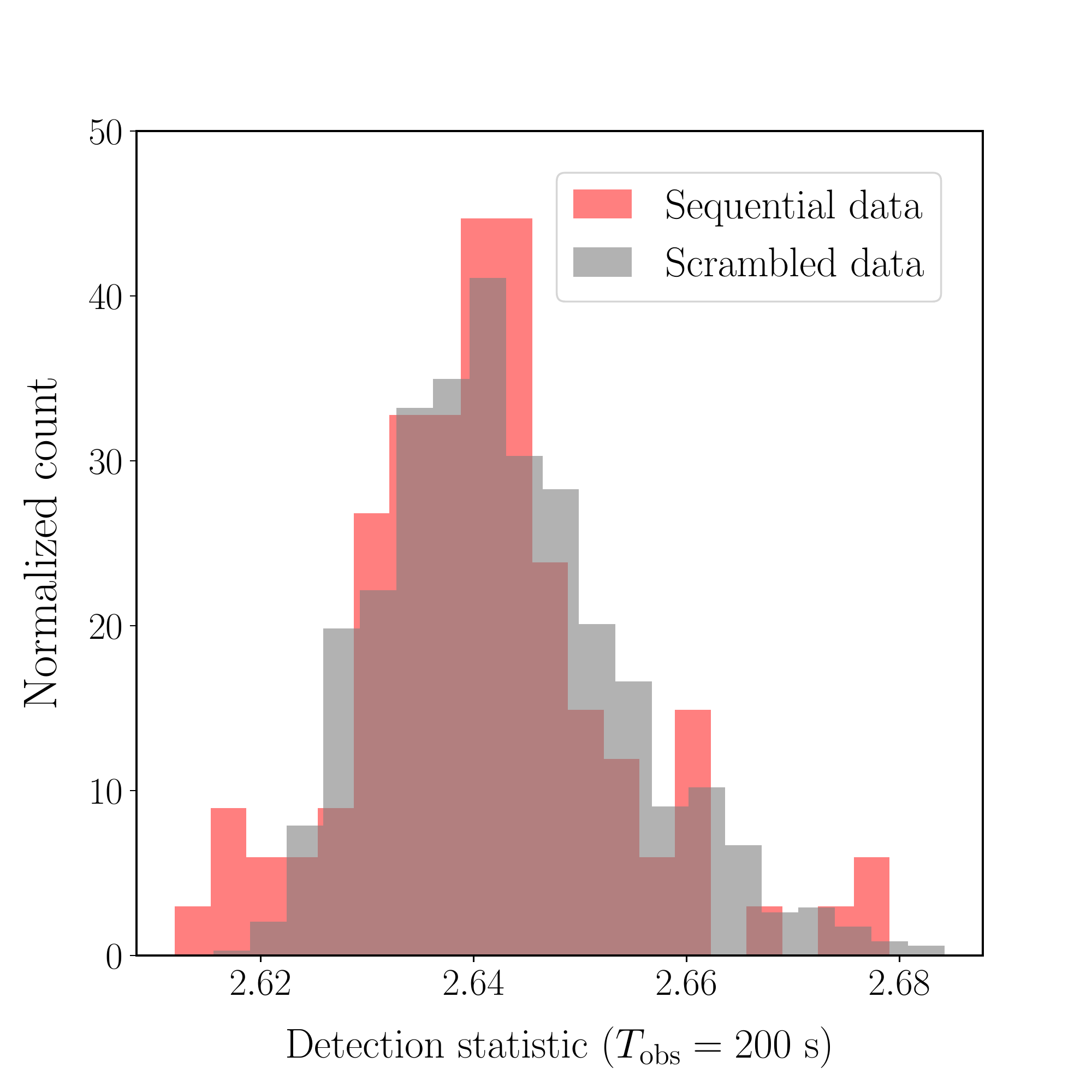}}	
	\subfigure[]{\label{fig:background_seq_600}
		\includegraphics[width=\columnwidth]{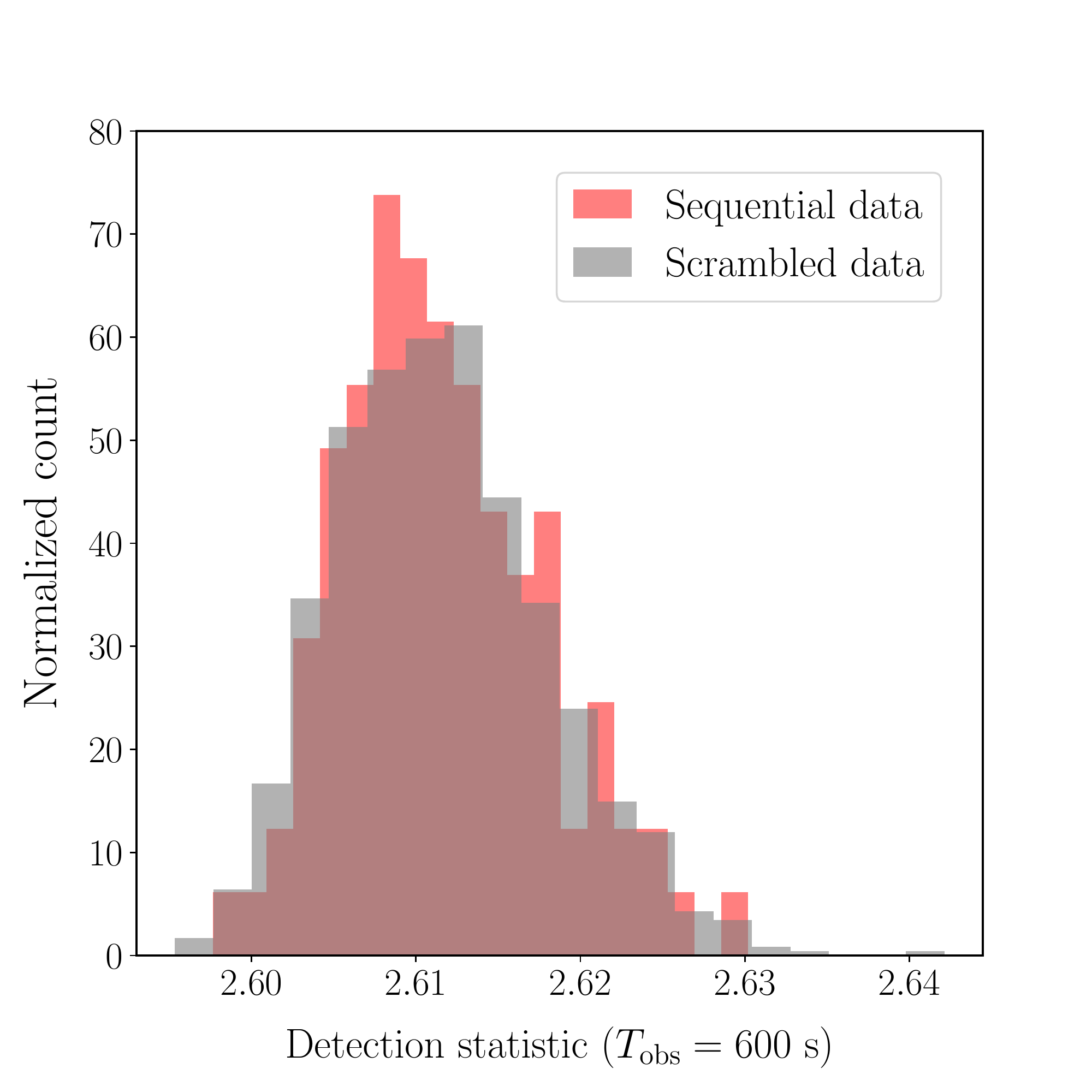}}
	\caption[]{Noise-only distribution of the detection statistic $\mathcal{P}$ from 100 sequential data samples (red) and 1000 scrambled data samples (gray) for (a) $T_{\rm obs} = 200$\,s and (b) $T_{\rm obs} = 600$\,s.}
	\label{fig:background_sample_seq}
\end{figure*}

To verify that the noise-only distributions of $\mathcal{P}$ in scrambled SFT sequences and unscrambled (i.e., sequential) data collected close to the event are comparable, we draw samples of sequential data from an observing period of six hours on the same day before the event, for $T_{\rm obs} = 200$\,s and 600\,s (100 samples each). The resulting distributions of $\mathcal{P}$ are plotted in Figure~\ref{fig:background_sample_seq}. The red and gray histograms indicate the distributions of $\mathcal{P}$ obtained from 100 samples of sequential data and 1000 samples of scrambled data, respectively. They generally agree with each other. For both $T_{\rm obs} = 200$\,s and 600\,s, the Kolmogorov-Smirnov test does not reject the null hypothesis at 5\% significance level (i.e., there is no significant difference between the two distributions).  Table~\ref{tab:Pth_compare} lists $\mathcal{P}_{\rm th}$ ($\alpha_{\rm f} = 1\%$) obtained from sequential and scrambled data sets. For both $T_{\rm obs} = 200$\,s and 600\,s, the discrepancy in $\mathcal{P}_{\rm th}$ is $\lesssim 0.1\%$. We do not conduct the verification for all $T_{\rm obs}$ choices because drawing enough sequential samples for larger $T_{\rm obs}$ values requires a few days' observing period, over which the impact from the time-varying detector PSD is no longer negligible.

\begin{table}
	\centering
	\setlength{\tabcolsep}{6pt}
	\renewcommand\arraystretch{1.4}
	\begin{tabular}{rrr}
		\hline
		\hline
		$T_{\rm obs}$ (s)& $\mathcal{P}_{\rm th}$ (sequential data) &$\mathcal{P}_{\rm th}$ (scrambled data) \\
		\hline
		200 & 2.6783 & 2.6750 \\
		600 & 2.6280 & 2.6287 \\
		\hline
		\hline
	\end{tabular}
	\caption[]{Detection threshold $\mathcal{P}_{\rm th}$ obtained from sequential and scrambled data ($\alpha_{\rm f} = 1\%$).}
	\label{tab:Pth_compare}
\end{table}

\section{Noise-only distribution of $\mathcal{P}$}
\label{sec:background}

Figure~\ref{fig:background2} provides a full set of noise-only distributions of the detection statistics (gray histograms) for all $T_{\rm obs}$ values used in the search. Each panel corresponds to one $T_{\rm obs}$ value. Each distribution is obtained from 1000 noise realizations. The $\mathcal{P}$ values obtained from the search targeting GW170817 are shown as colored dashed lines. The color of each dashed line indicates the $t_{\rm wait}$ value used, as shown in the legend.
The threshold $\mathcal{P}_{\rm th}$ ($\alpha_{\rm f} = 1\%$) is indicated by the black solid line. We have \mbox{$\mathcal{P} < \mathcal{P}_{\rm th}$} in all panels, indicating that the search results are consistent with the noise background.

\begin{figure*}
	\centering
	\subfigure{\scalebox{0.29}{\includegraphics{hist-200.pdf}}}
	\subfigure{\scalebox{0.29}{\includegraphics{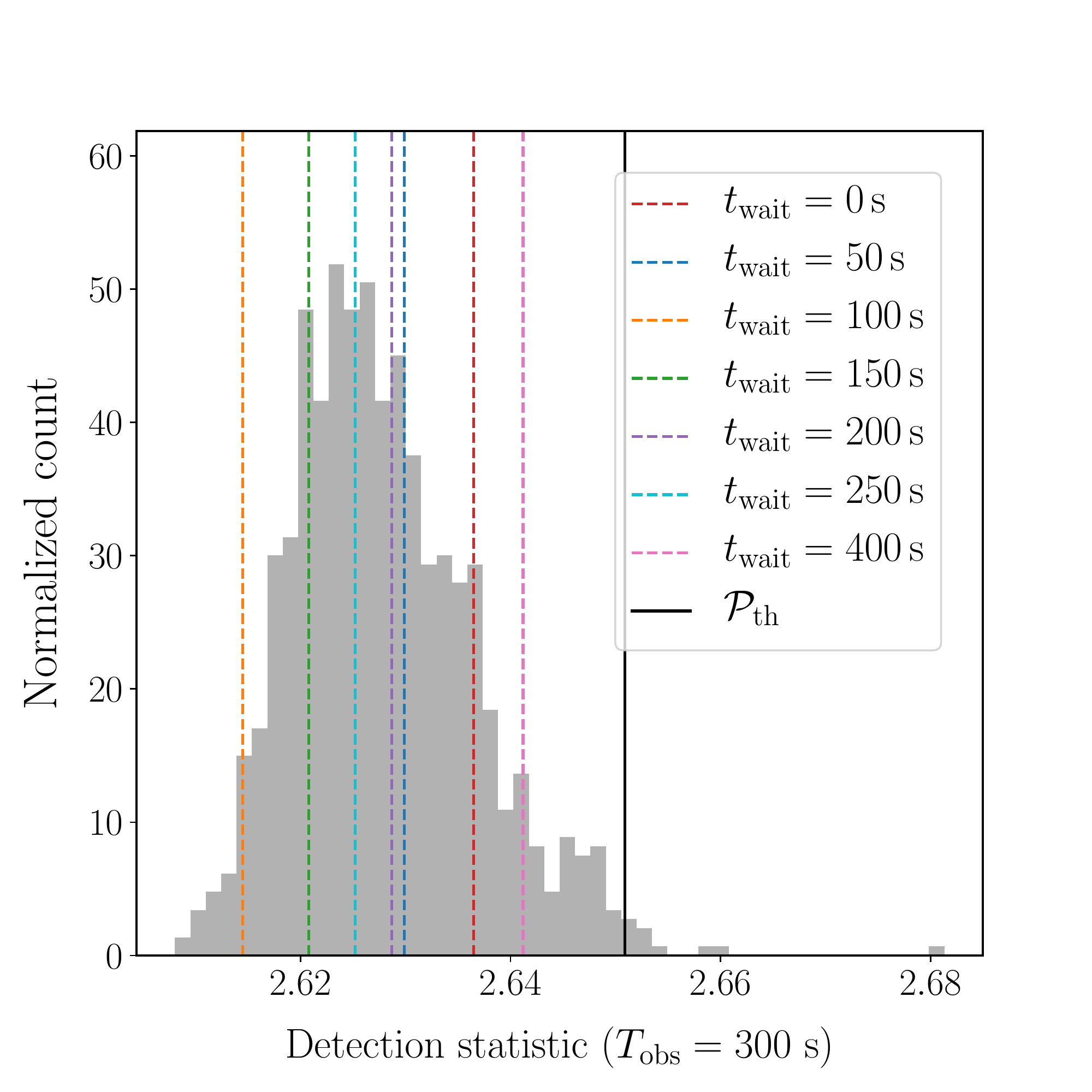}}}
	\subfigure{\scalebox{0.29}{\includegraphics{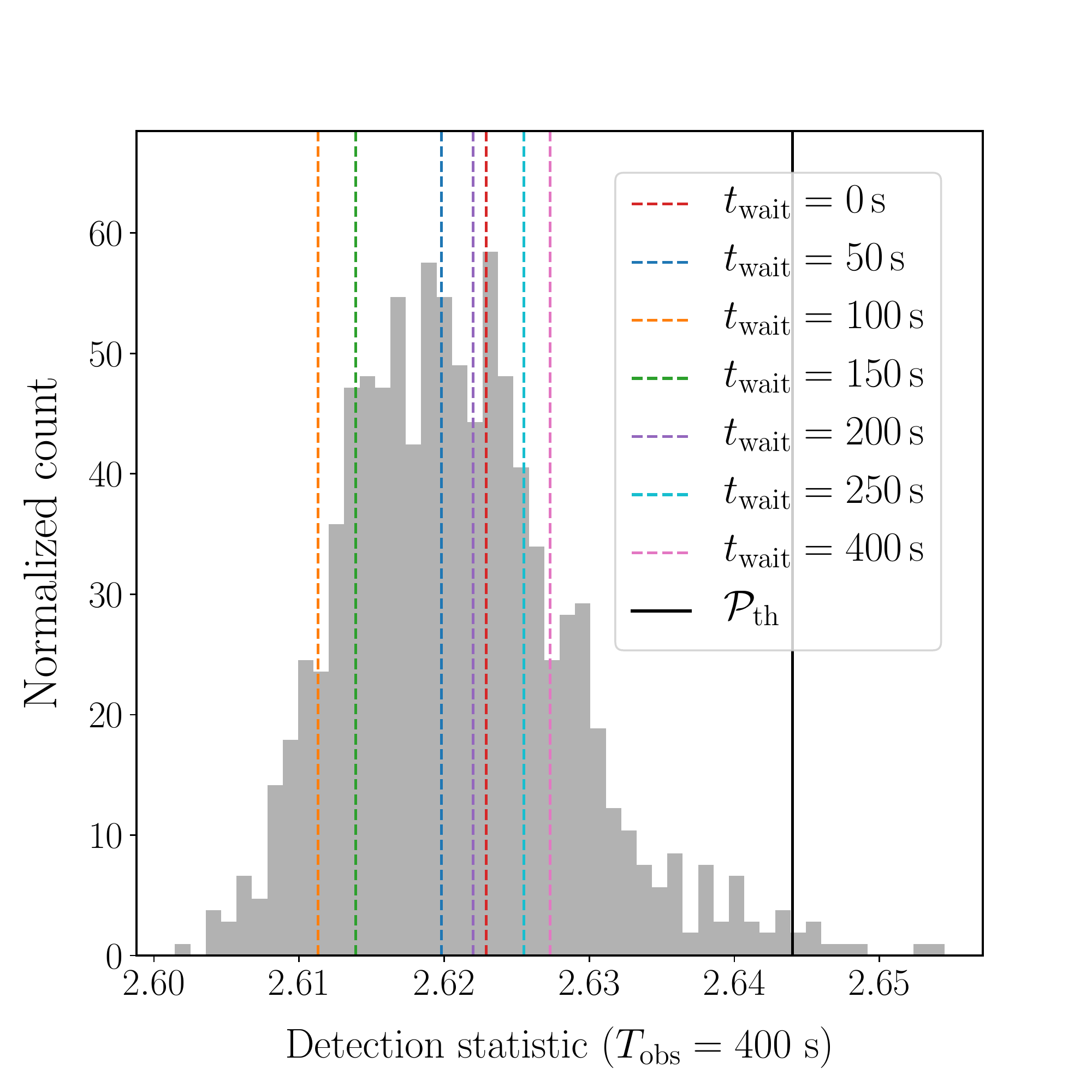}}}
	\subfigure{\scalebox{0.29}{\includegraphics{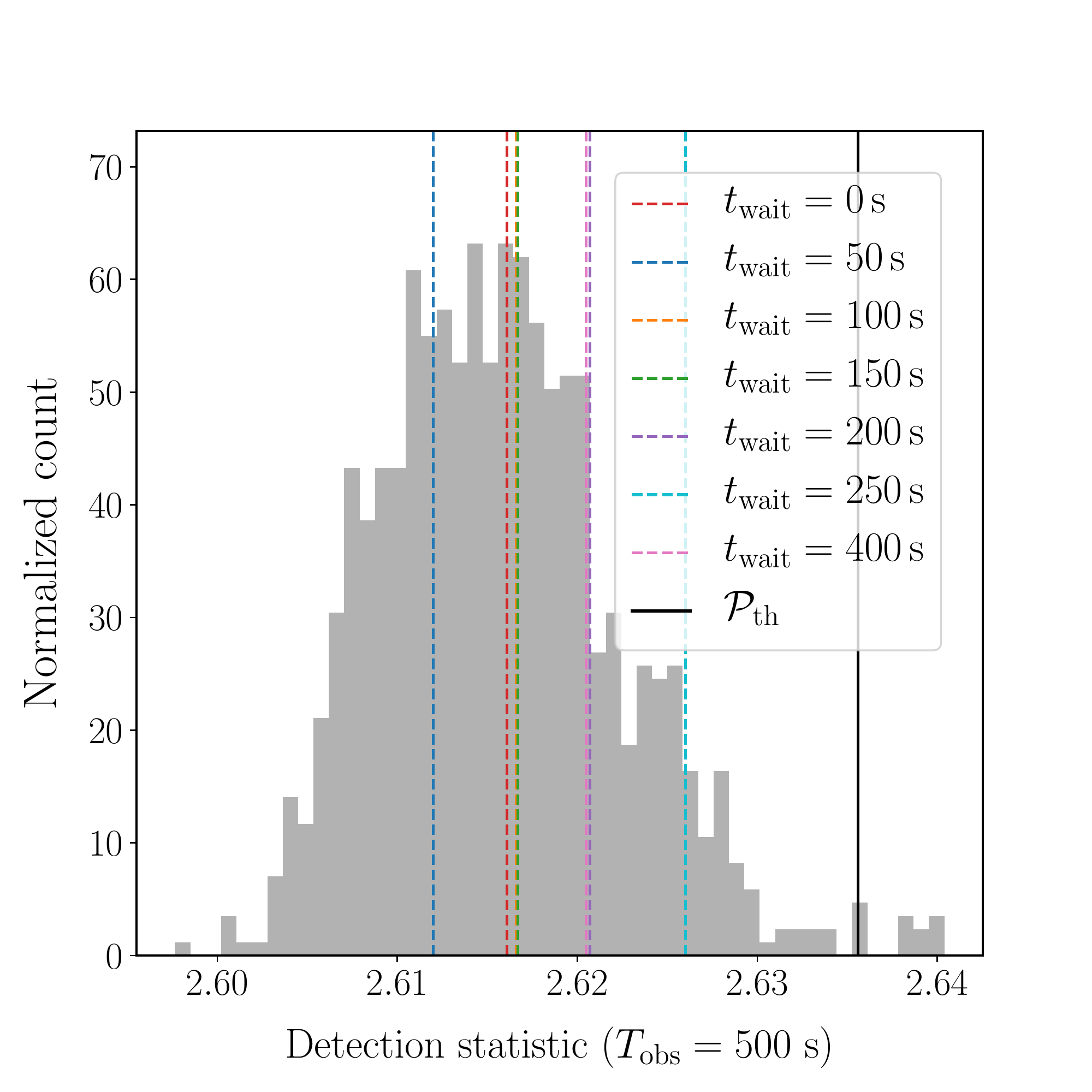}}}
	\subfigure{\scalebox{0.29}{\includegraphics{hist-600.pdf}}}
	\subfigure{\scalebox{0.29}{\includegraphics{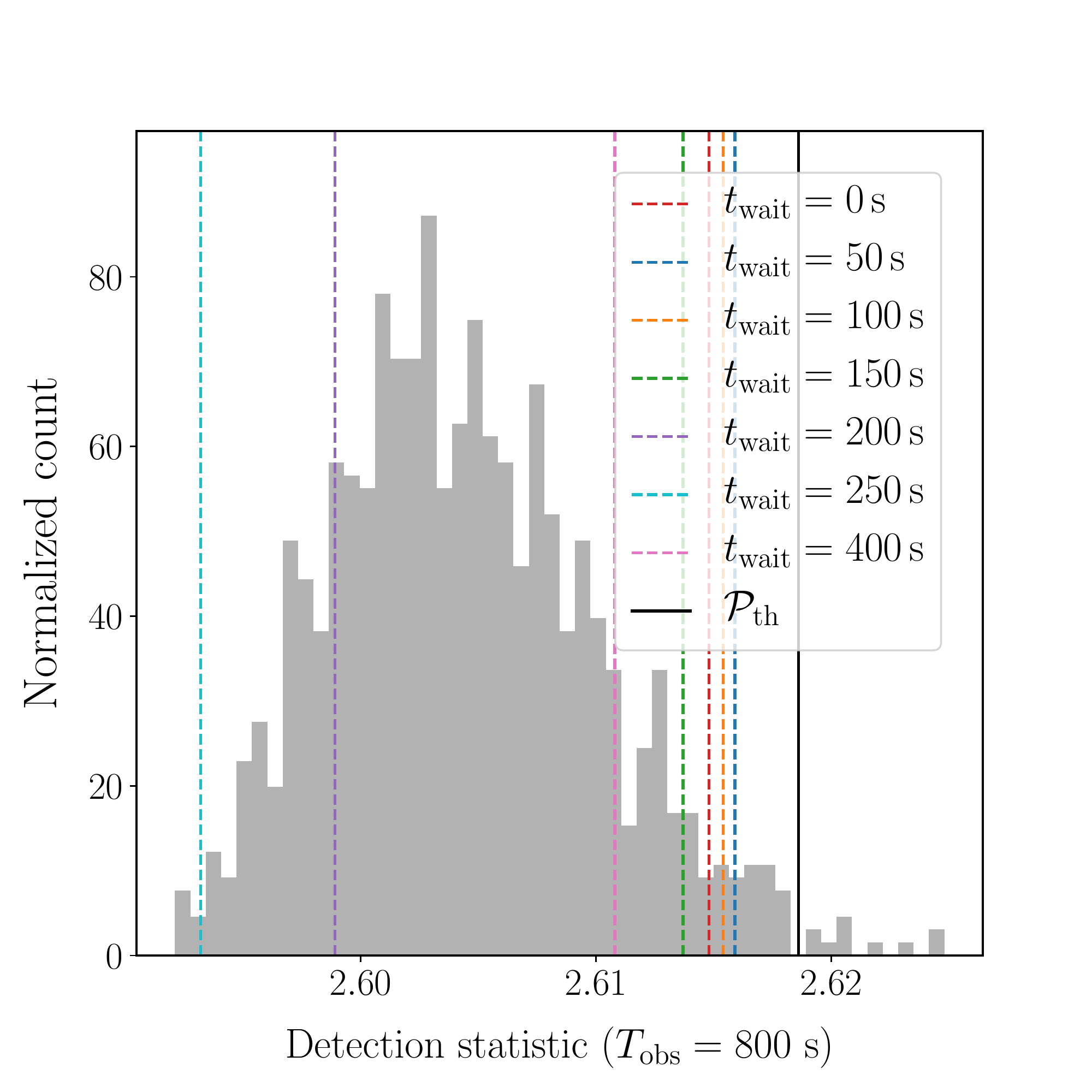}}}
	\subfigure{\scalebox{0.29}{\includegraphics{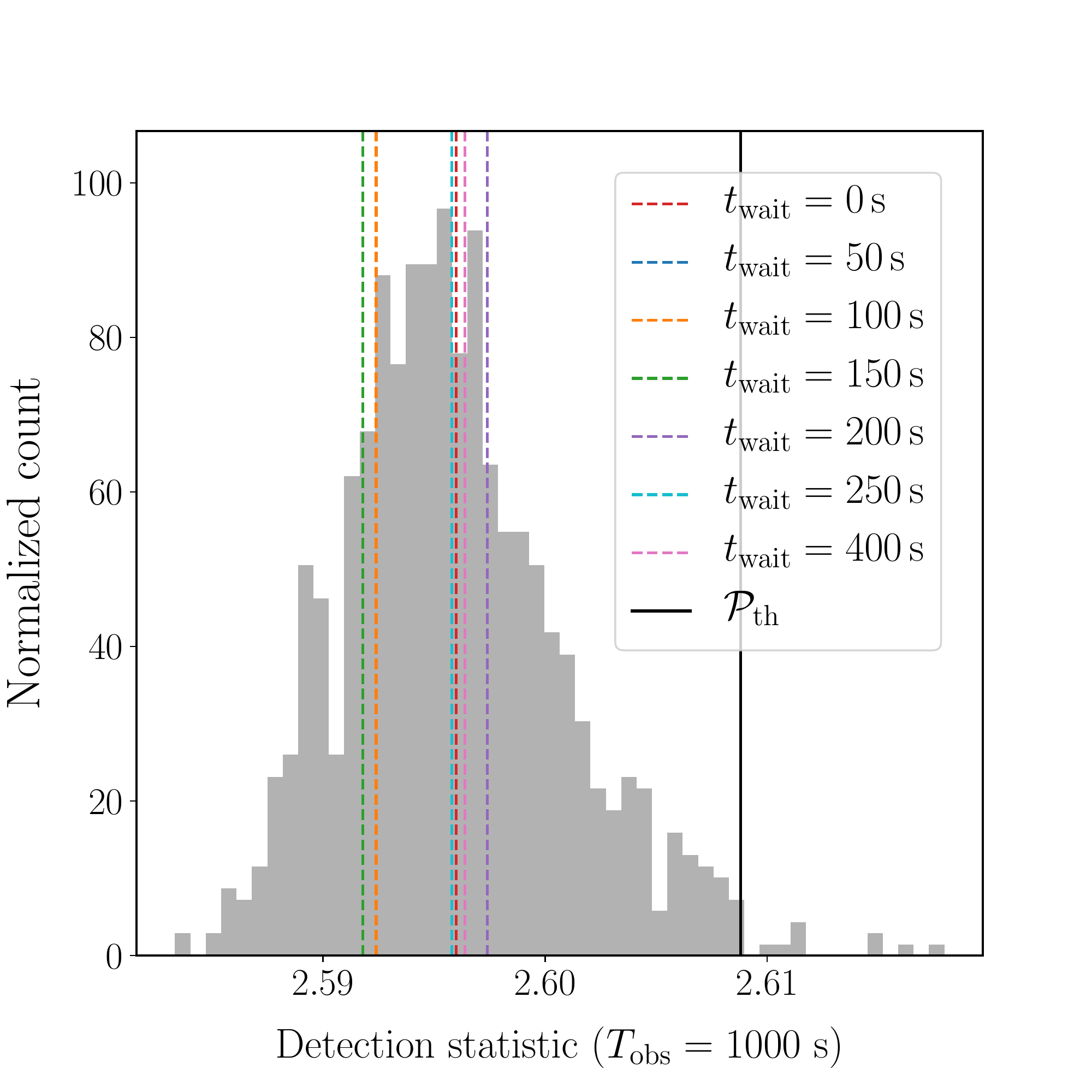}}}
	\subfigure{\scalebox{0.29}{\includegraphics{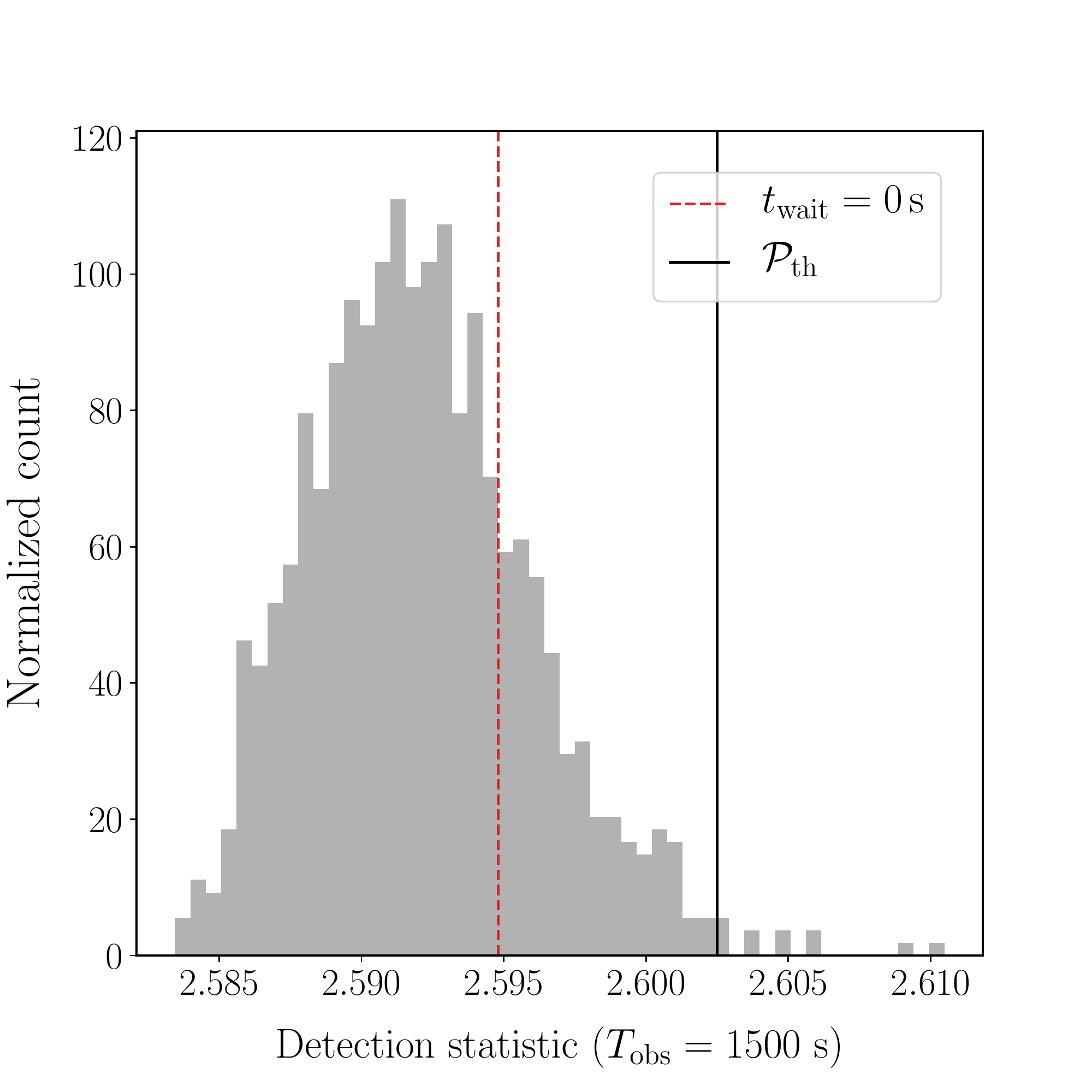}}}
	\subfigure{\scalebox{0.29}{\includegraphics{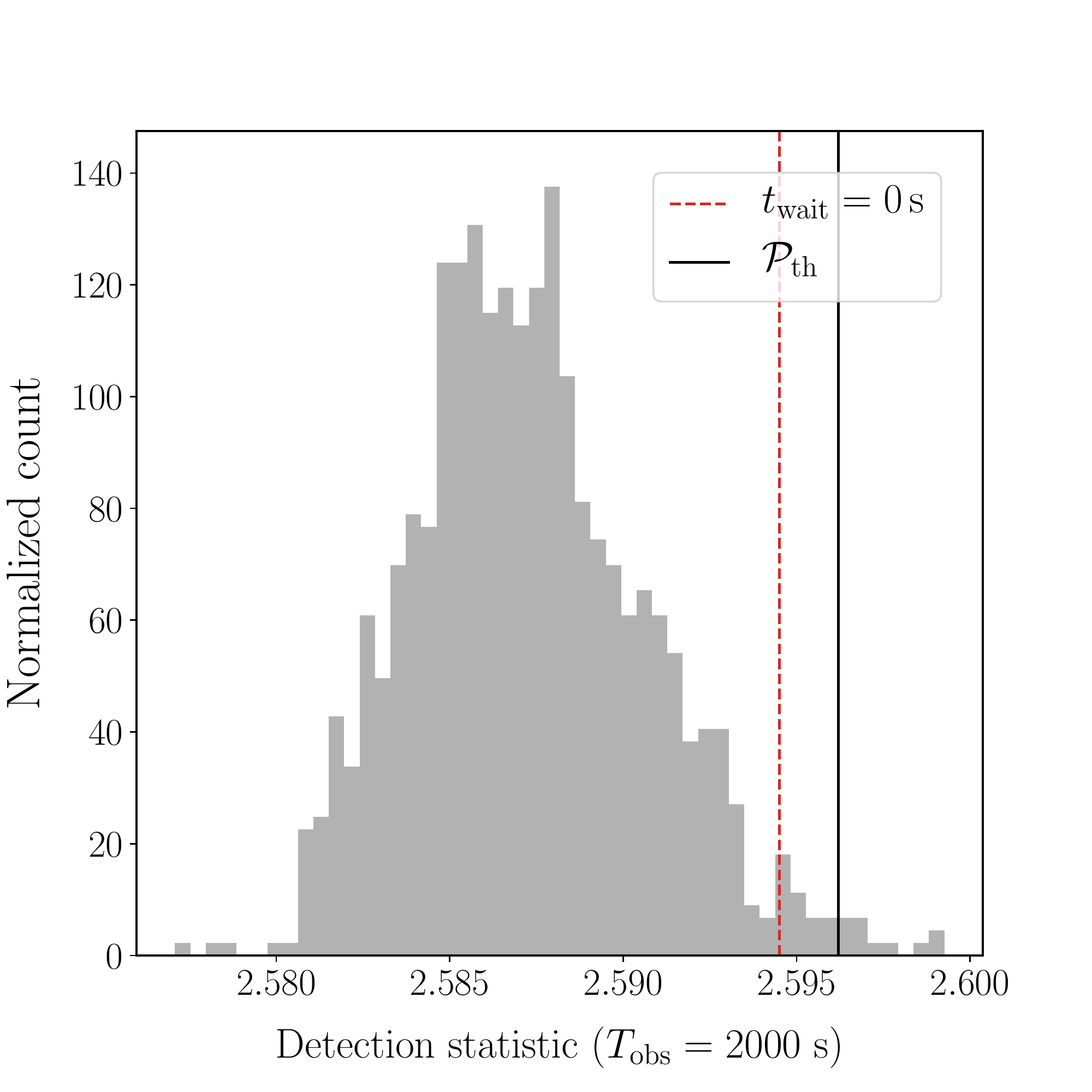}}}
	\subfigure{\scalebox{0.29}{\includegraphics{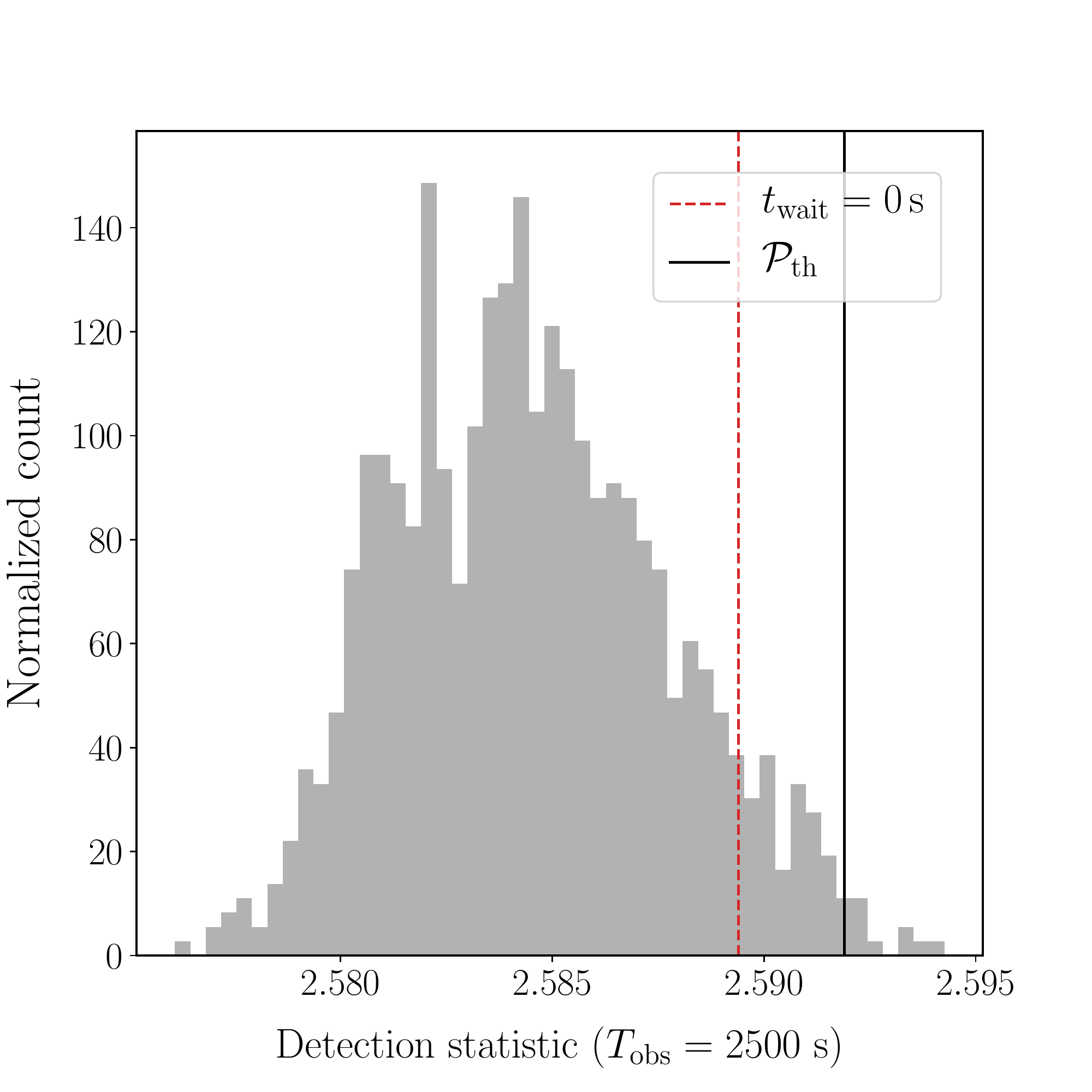}}}
	\subfigure{\scalebox{0.29}{\includegraphics{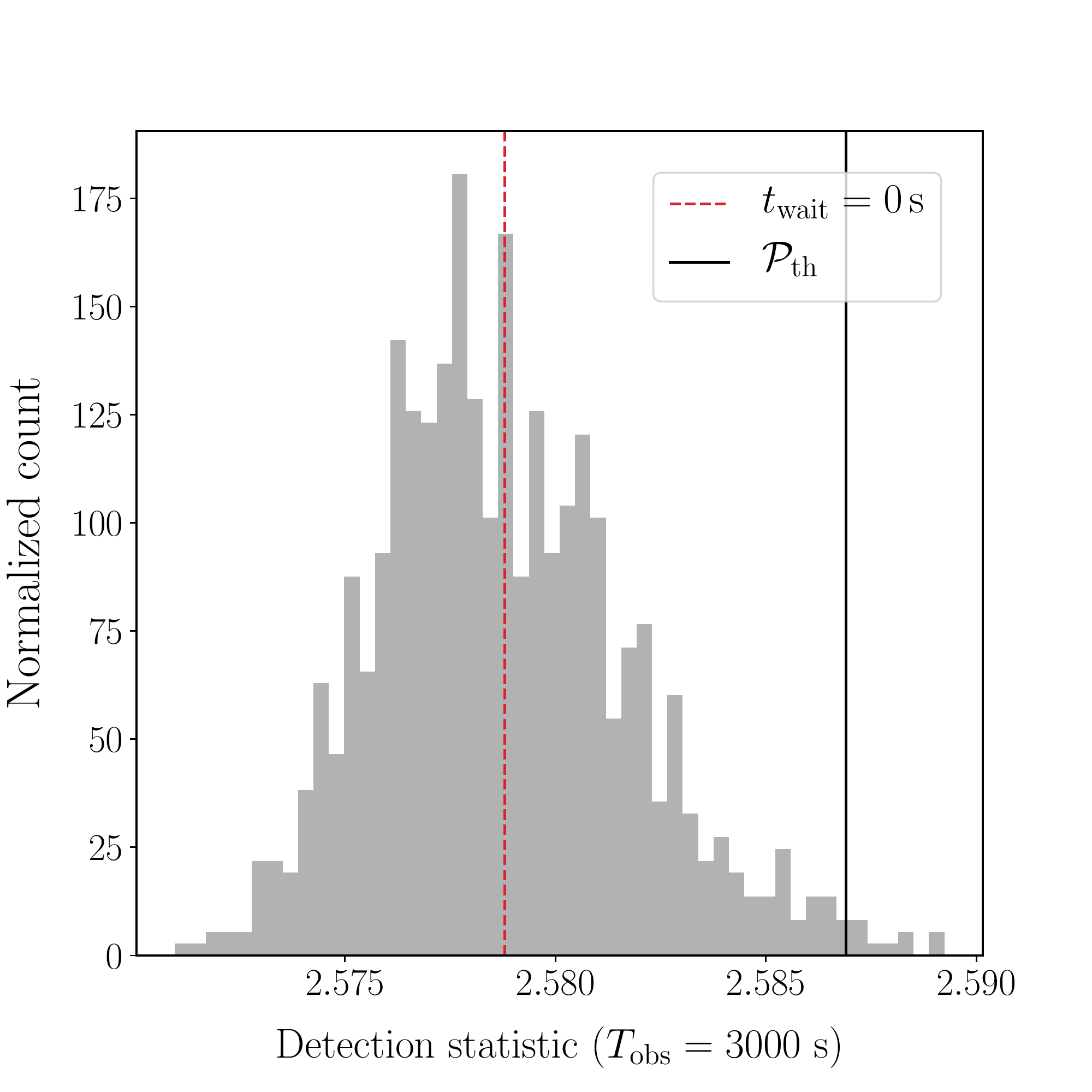}}}
	\subfigure{\scalebox{0.29}{\includegraphics{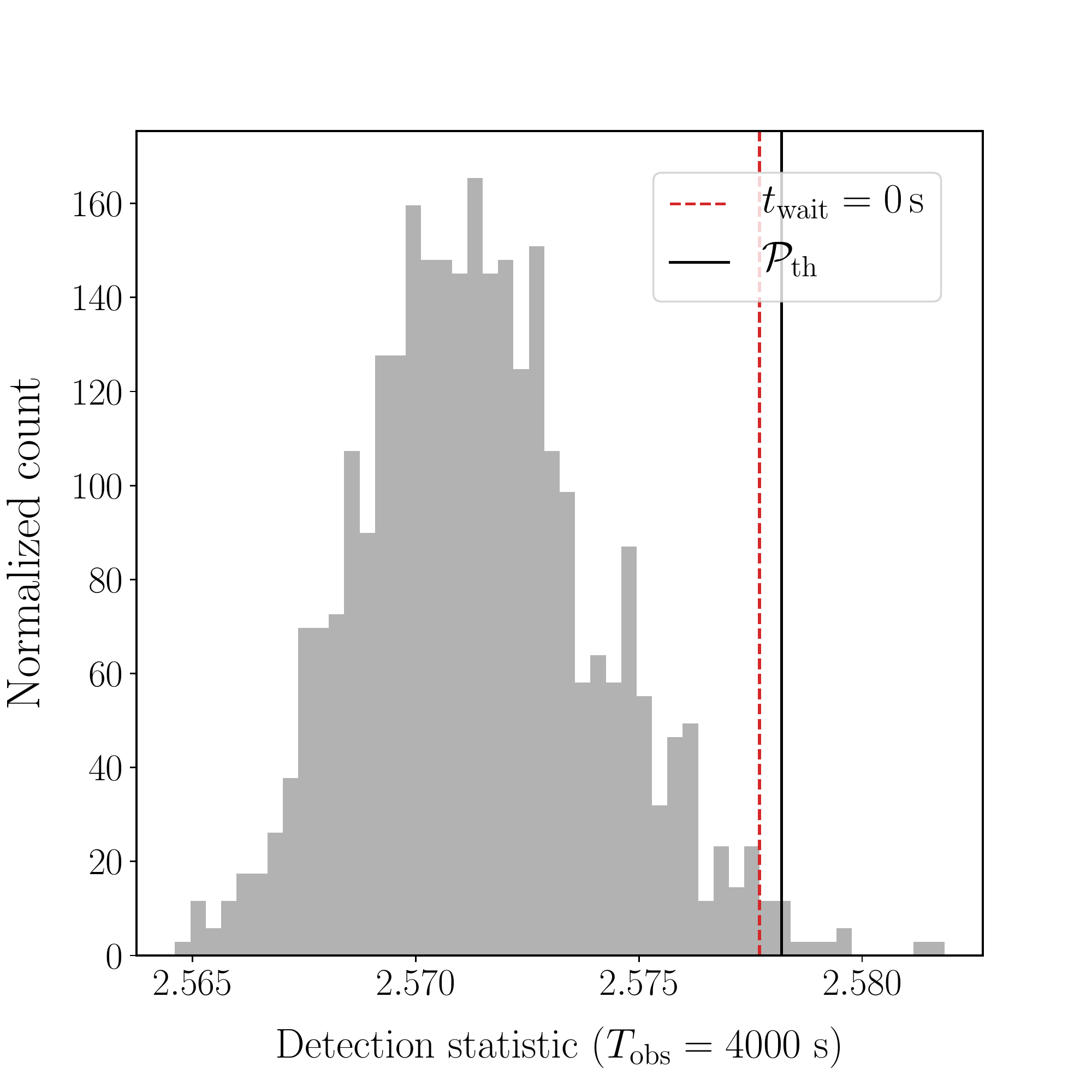}}}
	\label{fig:background}
\end{figure*}
\begin{figure*}
	\centering
	\subfigure{\scalebox{0.29}{\includegraphics{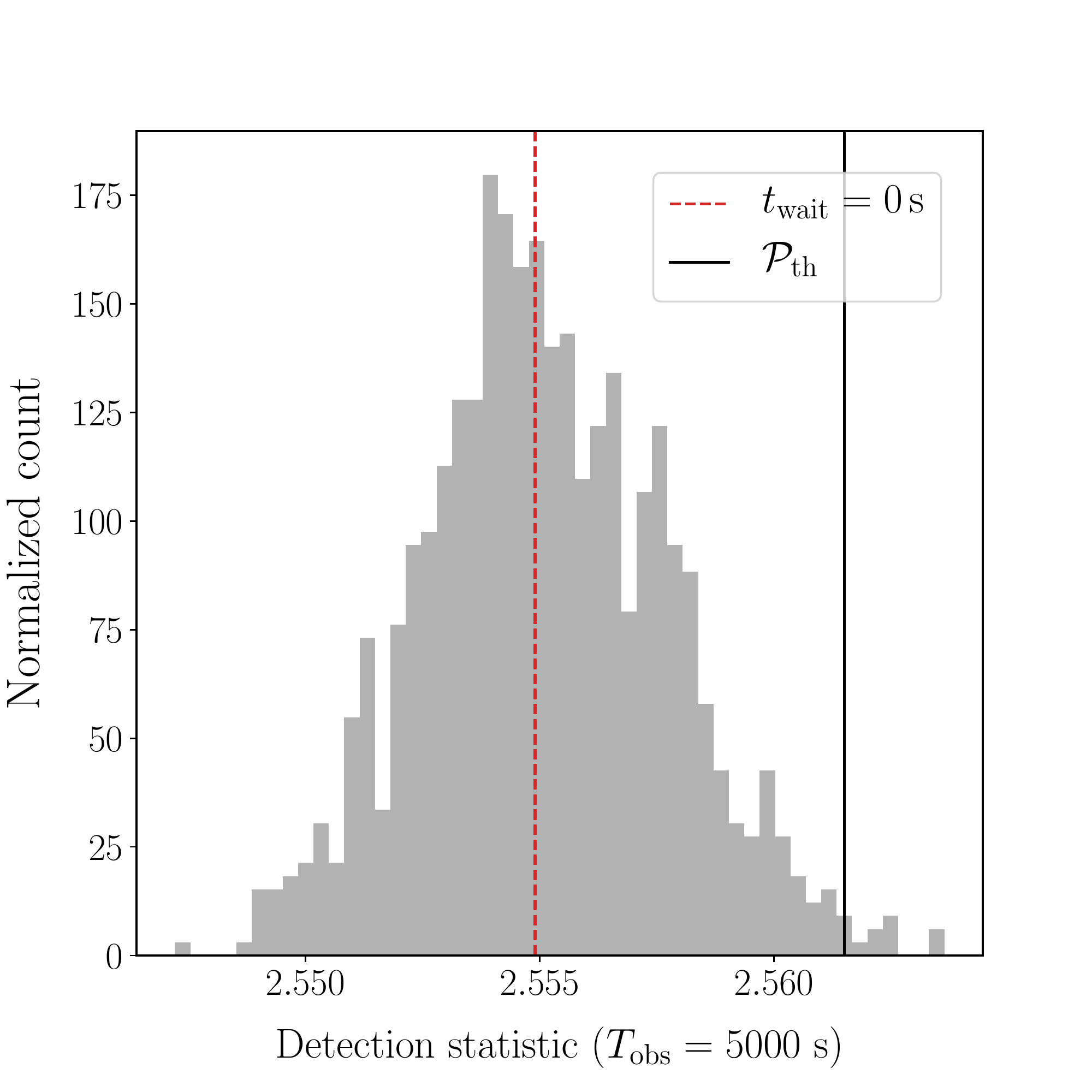}}}
	\subfigure{\scalebox{0.29}{\includegraphics{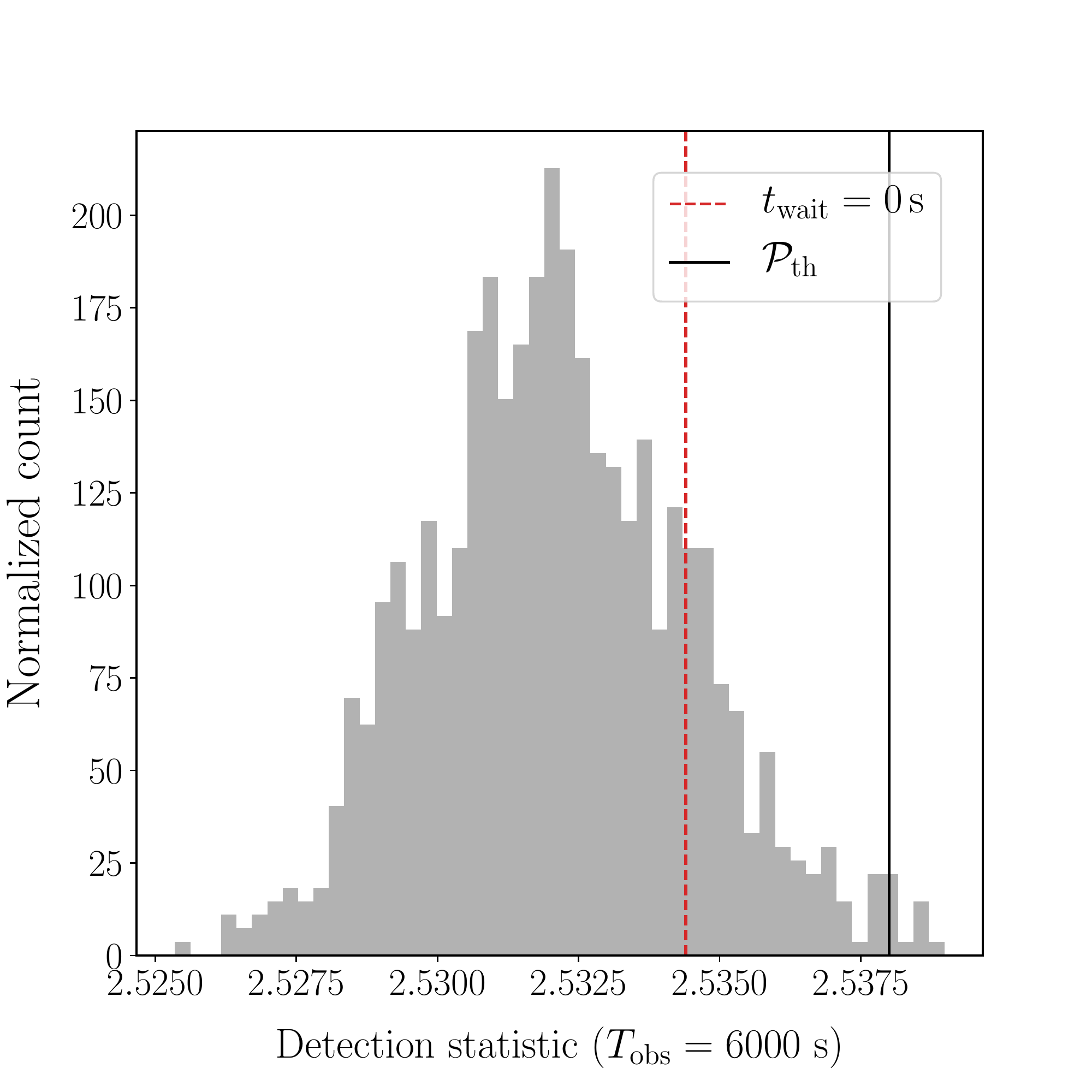}}}
	\subfigure{\scalebox{0.29}{\includegraphics{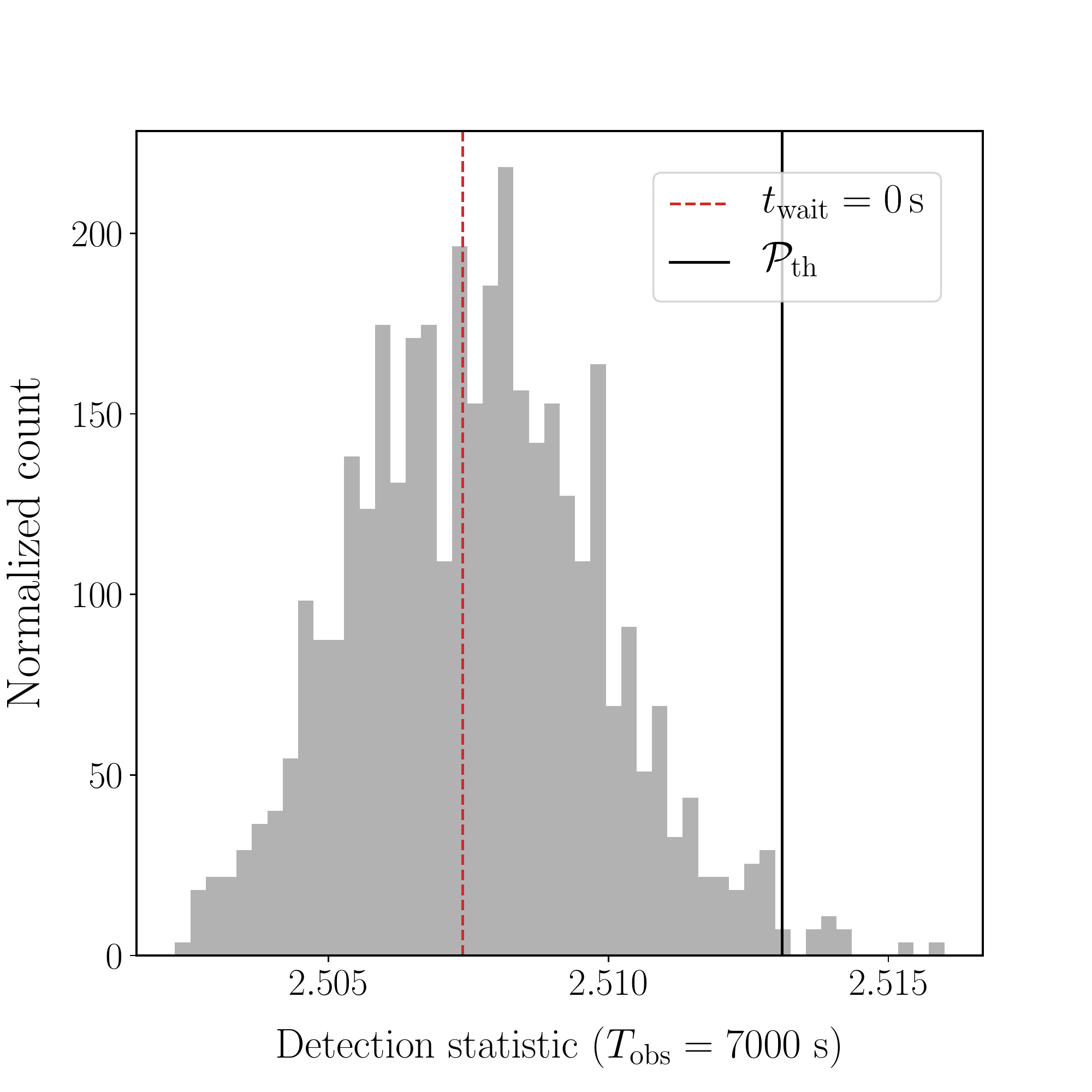}}}
	\subfigure{\scalebox{0.29}{\includegraphics{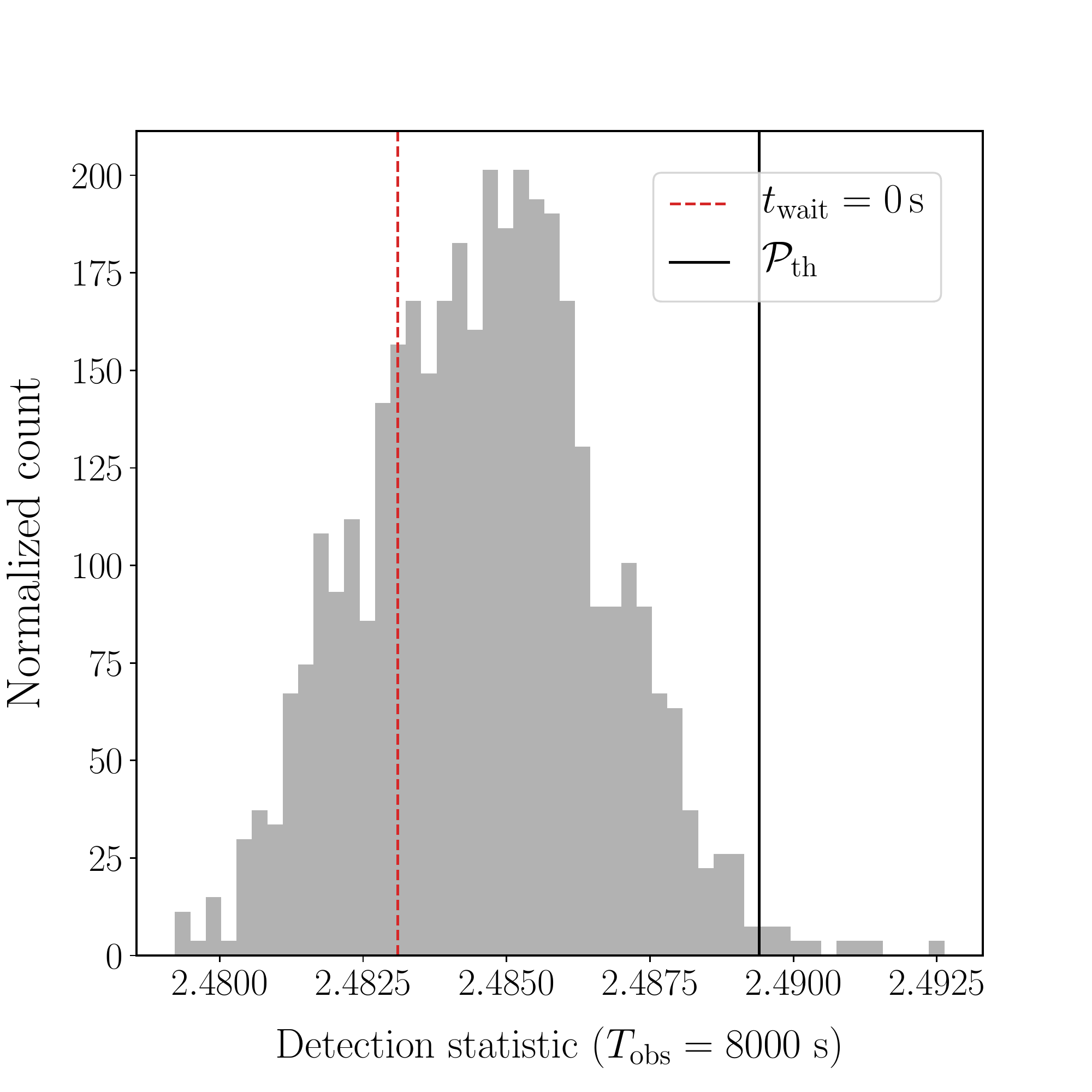}}}
	\subfigure{\scalebox{0.29}{\includegraphics{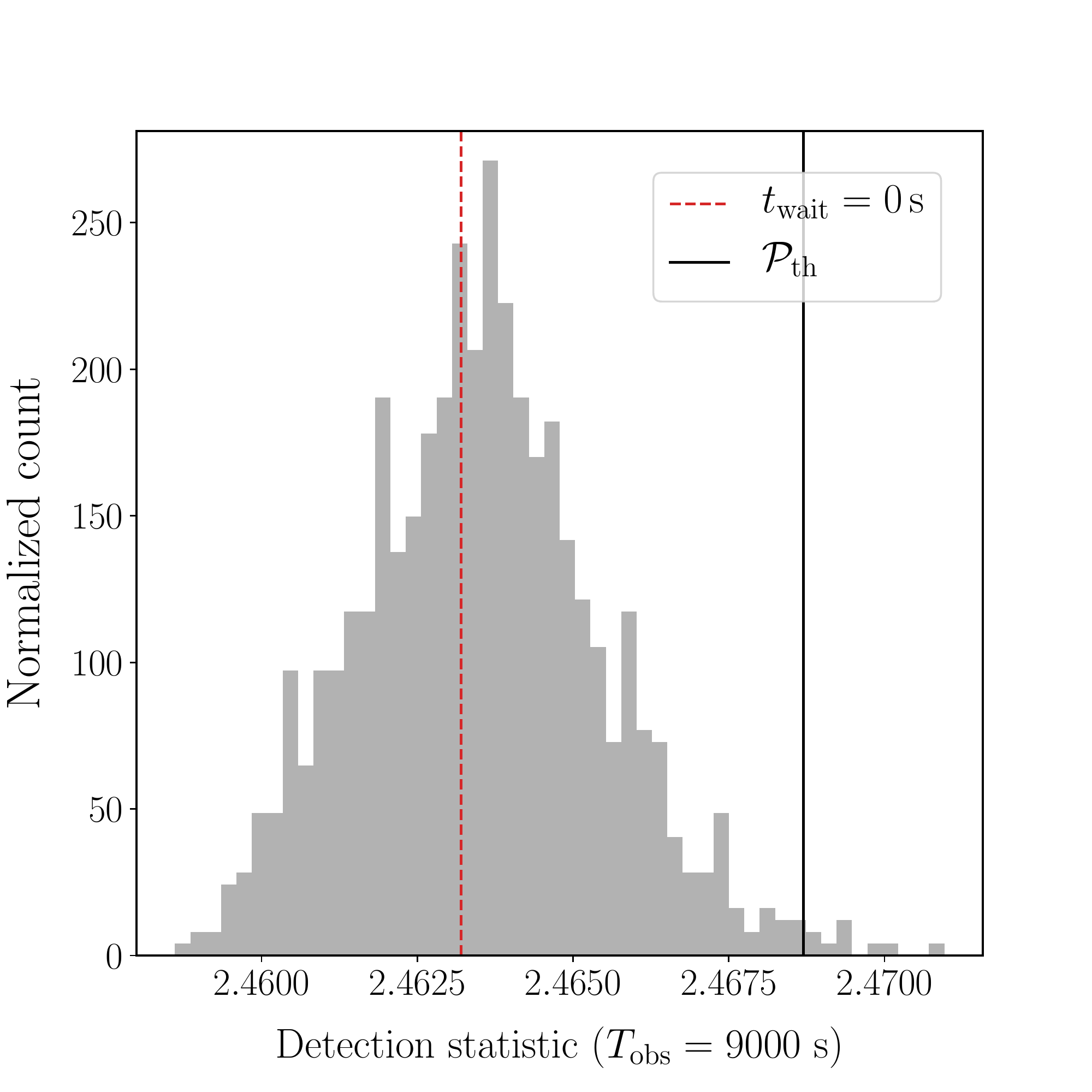}}}
	\subfigure{\scalebox{0.29}{\includegraphics{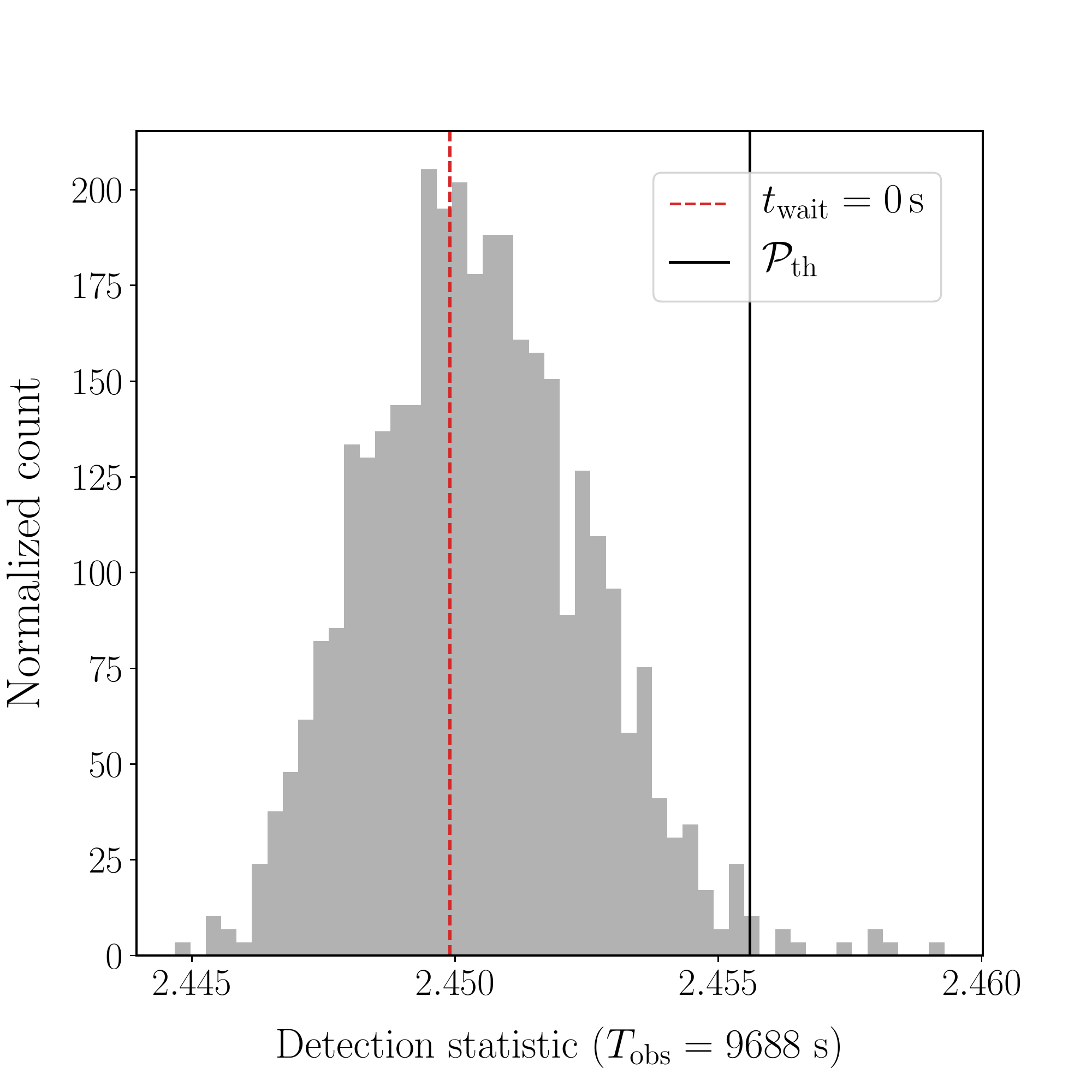}}}
	\caption[]{Noise-only distribution of the detection statistics (gray histogram) and $\mathcal{P}$ values obtained from the search targeting GW170817 (colored dashed lines), using various $t_{\rm wait}$ values. Each panel corresponds to one choice of $T_{\rm obs}$. The black solid line indicates the threshold $\mathcal{P}_{\rm th}$ ($\alpha_{\rm f} = 1\%$). The noise-only distribution is obtained from 1000 noise realizations for each panel.}
	\label{fig:background2}
\end{figure*}

\FloatBarrier

\end{document}